\documentclass[journal=jacsat , manuscript=article , layout=twocolumn ]{achemso}
\usepackage[version=3]{mhchem} 
\usepackage{geometry} 
\usepackage{graphicx}
\usepackage{caption}
\usepackage{subcaption}
\usepackage{color}

\newcommand{\etal}{\textit{et al.}}
\newcommand{\abinitio}{\textit{ab initio }}
\newcommand{\eg}{\textit{e.g.}}

\author{Martin Mayo}
\affiliation[University of Cambridge]
{Theory of Condensed Matter Group, Cavendish Laboratory, University of Cambridge, J. J. Thomson Avenue,
Cambridge CB3 0HE, United Kingdom}
\author{Kent J. Griffith}
\affiliation[University of Cambridge]
{Department of Chemistry, University of Cambridge, Lensfield Road, Cambridge CB2 1EW, United Kingdom}
\author{Chris J. Pickard}
\affiliation[University of Cambridge]
{Department of Materials Science and Metallurgy, 27 Charles Babbage Rd, Cambridge CB3 0FS, United Kingdom}
\author{Andrew J. Morris}
\affiliation[University of Cambridge]
{Theory of Condensed Matter Group, Cavendish Laboratory, University of Cambridge, J. J. Thomson Avenue,
Cambridge CB3 0HE, United Kingdom}
\email{ajm255@cam.ac.uk}

\title{\textit{Ab initio} study of phosphorus anodes for lithium and sodium-ion batteries}

\begin{document}

\begin{abstract}

Phosphorus has received recent attention in the context of
high-capacity and high-rate anodes for lithium and sodium-ion
batteries.
Here, we present a first principles structure prediction study combined with NMR calculations which gives us insights into its lithiation/sodiation process.
We report a variety of new phases phases found by AIRSS and the atomic species swapping methods. 
Of particular interest, a stable Na$_5$P$_4$--C2/m  structure and locally stable structures found less than 10 meV/f.u. from the convex hull, such as Li$_4$P$_3$--P2$_1$2$_1$2$_1$, NaP$_5$--Pnma and Na$_4$P$_3$--Cmcm. 
The mechanical stability of Na$_5$P$_4$--C2/m and  Li$_4$P$_3$--P2$_1$2$_1$2$_1$ has been studied by first principles phonon calculations .  
We have calculated average voltages which suggests that black phosphorus (BP) can be considered as a safe anode in lithium-ion batteries due to its high lithium insertion voltage, 1.5 V; moreover, BP exhibits a relatively low theoretical volume expansion compared with other intercalation anodes, 216\% ($\Delta V/V$).
We identify that specific ranges in the calculated shielding can be
associated with specific ionic arrangements, results which  play an
important role in the interpretation of NMR spectroscopy
experiments.
Since the lithium-phosphides are found to be insulating even at high
lithium concentrations we show that Li-P-doped phases with aluminium
have electronic states at the Fermi level suggesting that using
aluminium as a dopant can improve the electrochemical performance of P
anodes.
\end{abstract}

\section{Introduction}
Owing to their relatively high specific energy and capacity, Li-ion batteries (LIBs) are the energy source of choice for portable electronic devices \cite{Chu2012}. 
Despite the vast technological advances made since the first commercial LIB was released by Sony in 1991, the specific energy of commercial LIBs is limited to approximately 250 $\textrm{Whkg}^{-1}$, which is half the estimated need for a family  car to travel 300 miles without recharge \cite{LIBbook}. 
The demand of higher specific energies and capacities motivates the study of novel materials for the next generation of LIBs. 
In almost all conventional LIBs available for commercial use the cathode is typically a transition layered metal oxide, $\textrm{LiMO}_2$, with M=Co, Ni, Mn, etc,  and the anode material is graphite \cite{LIBbook,Nitta2014,Kim2011}. 
Intercalation electrodes experience slight changes during charge and discharge, \eg, less than 7\% volume change in C negative electrodes \cite{Nitta2014},  leading to a high capability of retaining their capacity over charge/discharge cycles. 
However, these electrodes suffer from low specific capacity due to  the limited intercalation sites available for Li ions  in the host lattice \cite{McDowell2013}, \eg, 372 $\textrm{mAhg}^{-1}$ for graphite. In order to overcome the capacity limitation of intercalation anodes,  it has been suggested to use different alloys of lithium as LIB anodes \cite{Kasavajjula2007,Zhang2011,Szczech2011,Qian2012,McDowell2013,Nitta2014}. 
A wide range of materials have been studied for this purpose, such as group IV and V elements, magnesium, aluminium and gallium among others \cite{Zhang2011,Nitta2014}. Alloy materials  can achieve 2-10 times higher capacity  compared to graphite anodes, where the highest capacity is achieved by silicon, 3579 $\textrm{mAhg}^{-1}$ \cite{Obrovac2004}. However, alloys tend to undergo relatively large structural changes under lithiation \cite{Obrovac2004,Zhang2011,LIBbook,Nitta2014}, leading to a poor cycle life.

Due to the high abundance, low cost, and relatively uniform geographical distribution of Na, Na-ion batteries (NIBs) have received recent attention. 
Despite some disadvantages, such as its larger ionic radius (1.02 \AA\ compared to 0.76 \AA\ for Li) and the lower cell potential of most Na systems \cite{klein2013conversion}, NIBs are considered to be one of the most promising alternatives to meet large-scale electronic storage needs \cite{Hong2013}. 
In spite of  similarities between elemental Li and Na, Na systems  present significantly different kinetic and thermodynamic properties \cite{Qian2012}. 
The widely used graphite negative anode in LIBs is not successful for NIBs \cite{Stevens2001} due to a poor specific capacity and bad cyclability.  
The Si alloy suggested for LIBs is not suitable for NIBs as the Na concentration in Na-Si systems is limited to 50 \% \cite{Hong2013}, therefore most recent studies have  focused on Na-Sn and Na-Sb systems \cite{Hong2013}.      
 
Reaction of phosphorus with three Li or Na atoms produces  Li$_3$P \cite{Brauer1937} and Na$_3$P \cite{Dong2005} respectively; which corresponds to a large theoretical capacity of 2596 mAhg$^{-1}$ and theoretical volume expansion  $\Delta V / V$ of 216 \%  for Li$_3$P and 391 \% for Na$_3$P from black phosphorus (BP). 
Of the several known allotropes, black phosphorus, red phosphorus, and the recently synthesised phosphorene  \cite{li2014black} have been studied as candidates for LIB and NIB anodes \cite{kulish2015,zhao2014,sun2015phosphorene} . Recent experimental studies \cite{Park2007,Sun2012,Marino2011,marino2012,Qian2012,Qian2013,ramireddy2015} showed  that  the addition of carbon to the phosphorus anode leads to an improvement in the reversibility of charge/discharge processes, probably due to an enhancement in its electrical conductivity and mechanical stability. 
The  study conducted by Qian \etal \cite{Qian2012,Qian2013}, showed that  amorphous phosphorus/carbon nano composite anodes are capable of achieving relatively high storage capacities per total mass of composite,  2355 $\textrm{mAhg}^{-1}$ for LIBs and 1765 $\textrm{mAhg}^{-1}$ for NIBs,  good capacity retention after 100 cycles and high power capabilities at high charge/discharge rates.  
All the experimental studies agree that Li$_3$P and Na$_3$P are formed  during the discharge process; however, the formation of other phases during the lithiation/sodiation process remains unclear. In the case of LIBs, differential capacity plots suggest the formation of Li$_x$P phases, however the assignment of the XRD spectra can be challenging  \cite{Park2007,Qian2012,Qian2013,sun2015phosphorene}. In a study presented by Sun \etal \cite{Sun2012} it has been suggested, based on \emph{ex situ} XRD, that crystal phases of Li-P form at the end of the charge.    

During Li/Na insertion and extraction, anodes are expected to form non-equilibrium structures. Evidence of a metastable structure formation in the Li-P system has been reported by Park \etal \cite{Park2007}, where the authors suggested the formation of Li$_2$P phase during the first discharge based on an electrochemical study. 
In the case of the well studied Li-Si system, the lithiation induces an electrochemical solid phase amorphisation, where the crystalline Si is consumed to form a Li$_x$Si amorphous phase \cite{McDowell2013}; nevertheless, the equilibrium crystalline compounds are generally used as a first step in order to study the electromechanical process (See Ref. \citenum{Hong2013} and references therein). 

\emph{Ab initio} techniques have been shown to be successful in giving insights into a better understanding  of different processes occurring in an electrode \cite{Ogata2014}. 
From total energies, important properties of an electrode like voltage profiles and volume change can be estimated.  
In addition, NMR parameters can be calculated for certain systems offering  a  powerful method to understand the local structure  of the studied system as well as a  way of complementing experimental studies. 

In this work, we present an \abinitio study of Li-P and Na-P compounds. 
We first perform a structure prediction study combining  atomic species swapping along with  \abinitio random structure searching methods for the Li-P and Na-P systems. 
We report various new stable and metastable structures and suggest connections between lithium/sodium contents and expected ionic arrangements. 
Lithiation/sodiation processes are assessed by calculating average voltage profiles, electronic density of states  and NMR chemical shifts of the ground state phases, allowing us to  predict the local environment evolution of P under lithiation/sodiation.  
We conclude showing the effect of dopants on the electronic structure of Li-P compounds, where we propose doping the anode with aluminium in order to improve the anode performance.           

\section{Methods}

Structure prediction was performed using the \abinitio random structure searching method (AIRSS) \cite{Pickard2006}. 
For a given system, AIRSS initially generates random structures which are then relaxed to a local minima in the potential energy surface (PES) using DFT forces. 
By generating large numbers of relaxed structures it is possible to widely cover the PES of the system. Based on general physical principles and system-specific constraints, the search can be biased in a variety of sensible ways \cite{Pickard2011}.
The phase space explored by the AIRSS method was extended by relaxing experimentally obtained crystal structures. 
All combinations of \{Li,Na,K\}-\{N,P,As,Sb\}  crystal structures at different stoichiometries were obtained from the International Crystallographic Structure Database (ICSD). 
For each structure, the anions and cations were swapped to Li/Na and P respectively and then relaxed using DFT forces. 
The AIRSS + species swapping method been successfully used for Li-Si \cite{Morris2014}, Li-Ge \cite{Morris2014,Jung2015(MorrisGe)} and Li-S \cite{See2014(MorrisS)} systems. Furthermore, a study on point defects in silicon has been presented \cite{Morris2008,Morris2013} using the AIRSS method.

AIRSS calculations were undertaken using the CASTEP DFT plane-wave code \cite{CASTEP}. The gradient corrected Perdew Burke Ernzerhof (PBE) exchange-correlation functional \cite{PBE} was used in all the calculations presented in this work.  
The core electrons were described using Vanderbilt "ultrasoft"   pseudopotentials,  the Brillouin zone was sampled using a Monkhorst-Pack grid \cite{MP} with a k-point spacing finer than $2\pi\times0.05\ \textrm{\AA}^{-1}$ . The plane wave basis set was truncated at an energy cutoff value of 400 eV for Li-P and 500 eV for Na-P. 

The thermodynamical phase stability of a system was assessed by comparing the free energy of different phases. 
From the available DFT total energy of a given binary phase of elements A and B, $E\{\textrm{A}_x\textrm{B}_y\}$, it is possible to define a formation energy per atom,
\begin{equation}
\label{ }
E_f/\textrm{atom}=\frac {E\{\textrm{A}_x\textrm{B}_y\} -xE\{\textrm{A\}}-yE\{\textrm{B}\}}{x+y}.
\end{equation}
The formation energies of each structure were then plotted as function of the B element concentration, $C_B=\frac{y}{x+y}$, starting at $C_B=0$ and ending at $C_B=1$. A convex hull was constructed between the chemical potentials at $(C_B,E_f/\textrm{atom})=(0,0);(1,0)$ drawing a tie-line that joins the lowest energy structures, provided that it forms a convex function. 
This construction gives access to the 0 K stable structure as the second law of thermodynamics demands that the (free) energy per atom is a convex function of the relative concentrations of the atoms (see Figure \ref{fig1}).

Average voltages for the structures lying on the hull were calculated from the available DFT total energies. For two given phases on the hull, $\textrm{A}_{x_1}\textrm{B}$ and $\textrm{A}_{x_2}\textrm{B}$ with $x_2>x_1$, the following two phase reaction is assumed,
\begin{equation}
\label{ }
\textrm{A}_{x_1}\textrm{B} + (x_2-x_1)\textrm{A}\rightarrow \textrm{A}_{x_2}\textrm{B}.
\end{equation}
The voltage, V, is given by \cite{Aydinol1997},
\begin{align*}
\label{ }
V =& -\frac{\Delta G}{x_2-x_1} \approx -\frac{\Delta E}{x_2-x_1}  \\
   = &-\frac{E(\textrm{A}_{x_2}\textrm{B})-E(\textrm{A}_{x_1}\textrm{B})}{x_2-x_1} + E(\textrm{A}),
\end{align*}
where it is assumed that  the Gibbs energy can be approximated by  the internal energy, as the pV  and thermal energy contributions are small \cite{Aydinol1997}.

The low energy structures obtained by the AIRSS search were refined with higher accuracy using  a  k-points spacing finer than $2\pi\times0.03\textrm{\AA}^{-1}$ and an energy cut-off of 650 eV for Li-P and 800 eV for Na-P and more accurate pseudopotentials \footnote{Pseudopotentials generated by the CASTEP on-the-fly generator:\\ Li 1|1.2|10|15|20|10U:20(qc=6) \\ Na 2|1.3|1.3|1.0|16|19|21|20U:30U:21(qc=8)\\ P 3|1.8|2|4|6|30:31:32}. 
The structures obtained from the ICSD were relaxed with the same level of theory and the formation energies and voltages were obtained. 
The same level of accuracy was used to calculate the nuclear magnetic shielding of the structures on the convex hull employing the gauge-including-projector-augmented-wave (GIPAW) algorithm \cite{GIPAW} implemented in CASTEP and the electronic density of states. 
The latter were calculated with the OptaDOS code \cite{Morris2014b} using the  linear extrapolative scheme \cite{Pickard1999,Pickard2000}.  
Phonon dispersion curves were calculated using Density Functional Perturbation Theory in CASTEP \cite{refson2006}.
Norm-conserving pseudopotentials \footnote{Pseudopotentials generated by the CASTEP on-the-fly generator: \\Li 1|1.2|18|21|24|10N:20N(qc=8) \\ Na 2|1.5|20|23|26|20N:30N:21N(qc=8) \\P 1|1.6|6|7|8|30N:31L:32N \\ } were used, the Brillouin zone was sampled using a Monkhorst-Pack grid \cite{MP} with a k-point spacing finer than $2\pi\times0.03\ \textrm{\AA}^{-1}$  and the plane wave basis set was truncated at an energy cut-off of 1000 eV. The structures were fully relaxed at this level of accuracy. 
The NMR parameters and density of states of black P were calculated with the CASTEP semi empirical dispersion correction \cite{mcnellis2009azobenzene}, using the scheme of Grimme (G06) \cite{grimme2006semiempirical}.

\section{Results}

\subsection{Lithium phosphide} \label{Results_LiP}

\begin{figure*}
\begin{center}
\includegraphics[width=0.65\textwidth]{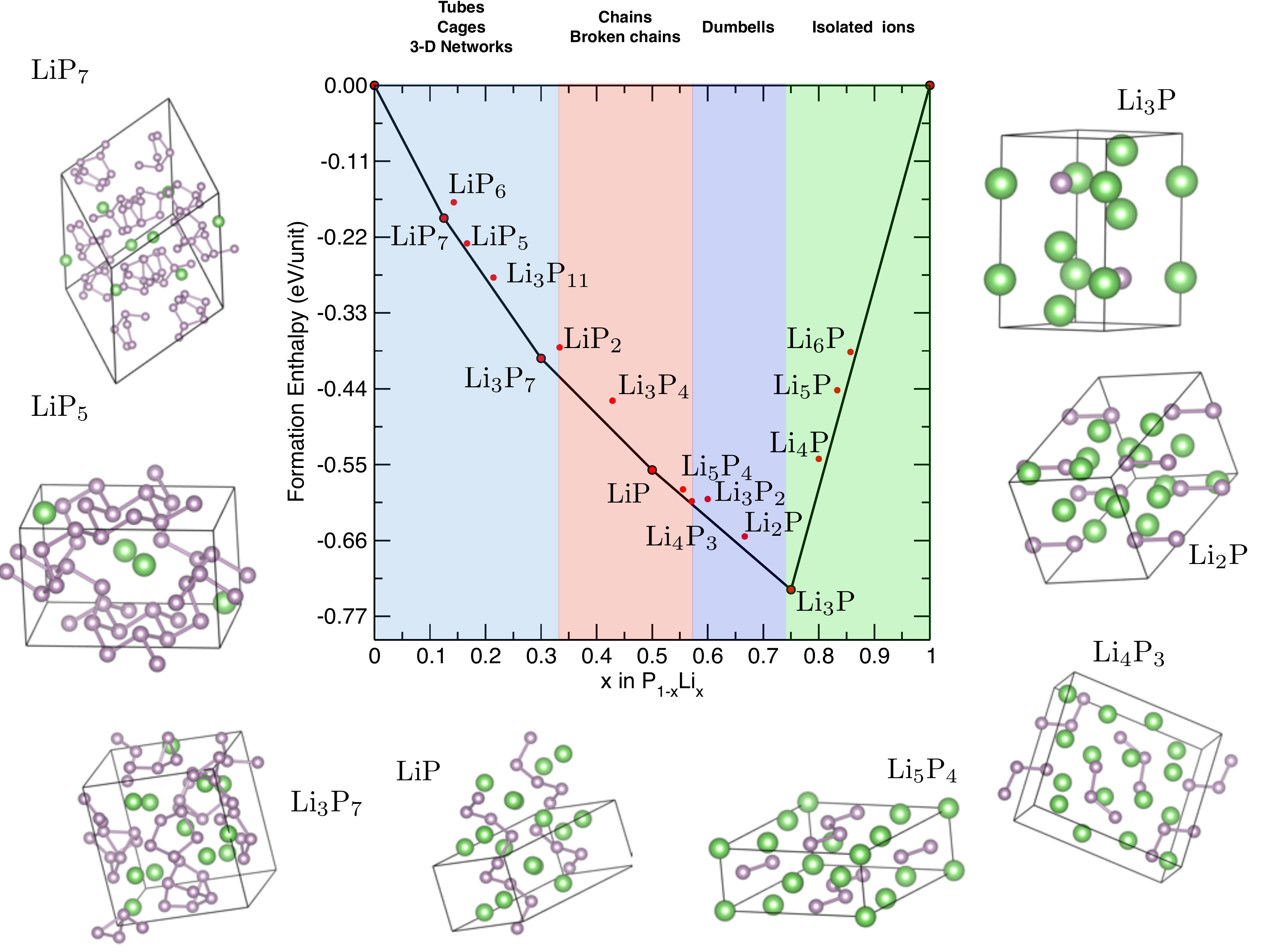}
\caption[Convex hull of Li-P system.]{Formation enthalpy per atom versus the fractional lithium concentration in the Li-P compound. The convex hull (tie-line) is constructed by joining the stable structures obtained by the searches. The convex hull has been divided in four main regions to guide the eye, highlighting the kind of ionic arrangement in each region. Selected structures are shown with green and purple spheres
denoting Li and P atoms, respectively, with the purple lines indicating
P--P bonds. For a full description of the phases, see Table \ref{LiPtable}. }
\label{fig1}
\end{center}
\end{figure*}

\begin{figure}
	\begin{center}
	\includegraphics[width=0.45\textwidth]{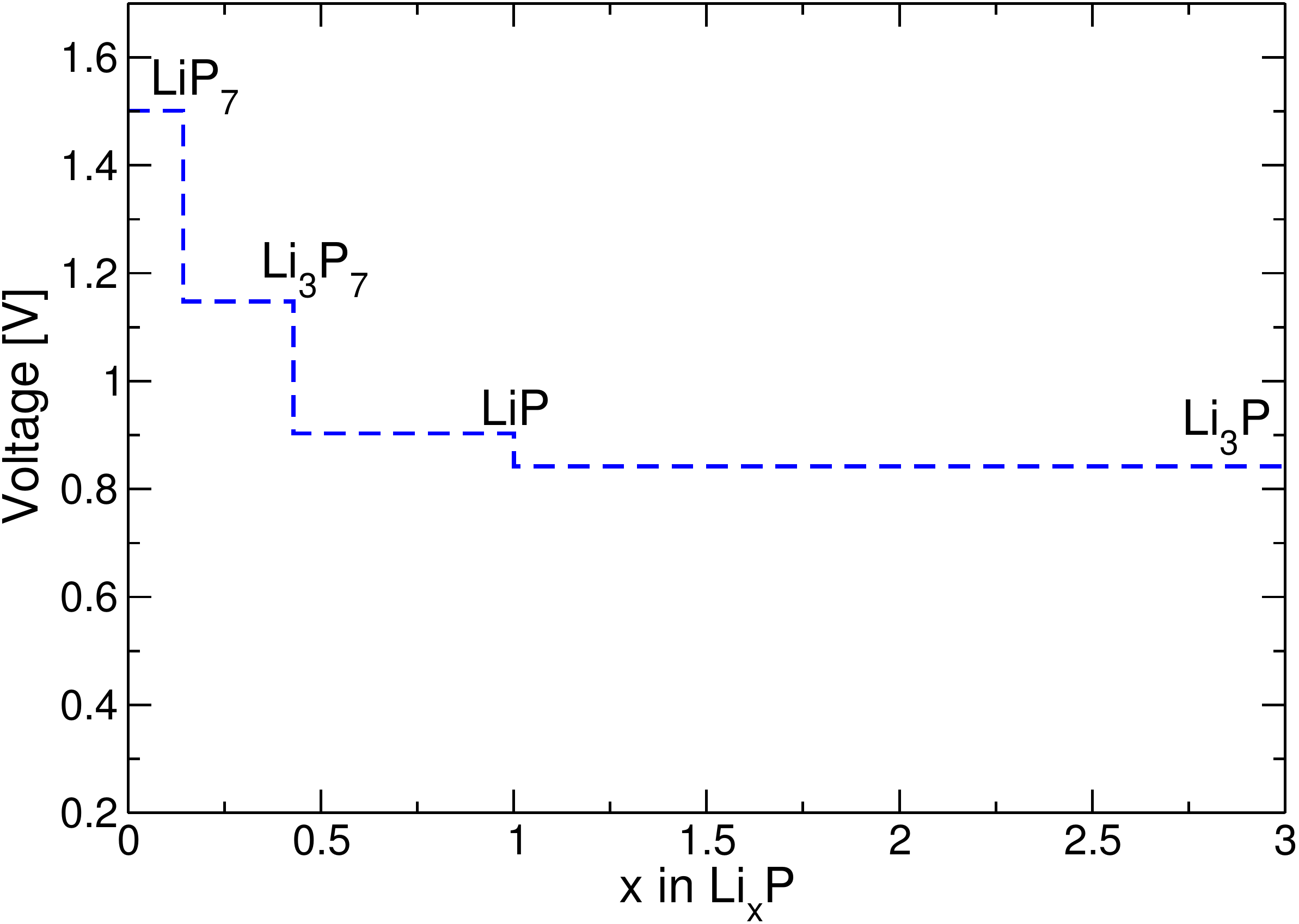}
	\caption[Voltage profile of Li-P system.]{Average voltages relative to lithium metal calculated for the structures found on the convex hull (Figure \ref{fig1}). }
	\label{fig3}
	\end{center}
\end{figure}

Figure \ref{fig1} shows the formation energy as a function of lithium concentration of the low-energy structures obtained by the search. 
The stable structures found on the convex hull, in increasing lithium concentration order, are black P--Cmca,  LiP$_7$--I4$_1$/acd \cite{Schnering1972}, Li$_3$P$_7$--P2$_1$2$_1$2$_1$ \cite{Honle1983},  LiP--P2$_1$/c  \cite{Honle1981}, Li$_3$P--P6$_3$/mmc  \cite{Dong2007} and Li--Im$\bar3$m.
A  novel DFT Li$_4$P$_3$--P2$_1$2$_1$2$_1$ phase is found 4 meV/f.u. above the convex hull, well within DFT accuracy. 
All the known Li-P phases are found on the convex hull, except for LiP$_5$--Pna2$_1$ \cite{Schnering1972} which is found 12 meV/f.u. from the convex tie-line in our 0 K DFT calculation.
The average voltage profile was calculated between pairs of proximate stable structures relative to Li metal. A plot of the average voltages as a function of Li concentration is presented in Figure \ref{fig3}.

Table \ref{LiPtable} summarises the structures presented in Figure \ref{fig1}. The  convex hull construction reveals new metastable structures, which are of importance when studying the lithiation process of the anode during cycling, as the anode is unlikely to reach thermodynamic equilibrium during charge and discharge. We identify that the structures can be categorised in four main regions according to their P ionic arrangement, as is highlighted in Figure \ref{fig1}.  As the Li concentration is increased, the structures change as following: tubes, cages and 3-D networks $\rightarrow$ chains and broken chains $\rightarrow$ P dumbbells $\rightarrow$ isolated P ions. 
\begin{table*} 
\centering
\caption[Li-P  phases found on the convex hull.]{Description of the experimental and predicted Li$_x$P phases. We indicate with a star ($\star$) the stable phases which are found on the convex hull. We identify four main regions with different ionic arrangements; for $0 \leq x \leq 0.5$ the structures show tubes, cages and 3-D networks composed of three and two P bonds, for $0.5 < x < 1.33$  P chains and broken chains,  for $1.33 < x \leq 2$ P dumbells and for concentrations larger than $x=2$ the structures are mainly composed of isolated P ions.  } \label{LiPtable}
\resizebox{\linewidth}{!}{
 \begin{tabular}{lccclc}
    \hline
    Stoichometry     & x in Li$_x$P & Distance from & Space            & Structure origin                                    & Description           \\
    ~                & ~            & the hull [eV/f.u.] & group            & ~                                                   & ~                     \\ \hline
    black-P  $^\star$          & 0            & ~             & Cmca             & ~                                                   & ~                     \\
    LiP$_7$ $^\star$     & 0.143        & ~             & I4$_1$/acd       & Known Li-P phase \cite{Schnering1972}               & P tubes               \\
    LiP$_6$          & 0.167        & 0.046         & R$\bar3$m        & AIRSS                                               & P 3-D network         \\
    LiP$_5$          & 0.2          & 0.012         & Pna2$_1$         & Known Li-P phase \cite{Schnering1972}               & P 3-D network         \\
    Li$_3$P$_{11}$   & 0.273        & 0.017         & Pbcn             & Swapping from  Na$_3$P$_{11}$ \cite{Wichelhaus1973} & P$_{11}$ cages        \\
    Li$_3$P$_7$ $^\star$ & 0.429        & ~             & P2$_1$2$_1$2$_1$ & Known Li-P phase \cite{Honle1983}                   & P$_{7}$ cages         \\
    LiP$_2$          & 0.5          & 0.041         & P2$_1$           & AIRSS                                               & Black P - like layers \\
    Li$_3$P$_4$      & 0.75         & 0.043         & C2/m             & AIRSS                                               & Chair-like chains     \\
    LiP   $^\star$       & 1            & ~             & P2$_1$/c         & Known Li-P phase \cite{Honle1981}                   & P helix               \\
    Li$_5$P$_4$      & 1.3          & 0.012         & C2/m             & Swapping from Na$_5$As$_4$ \cite{Ozisik2011}        & 4 P zig-zag chains    \\
    Li$_4$P$_3$      & 1.333        & 0.006         & P2$_1$2$_1$2$_1$             & AIRSS                                               & 3 P zig-zag chains    \\
    Li$_3$P$_2$      & 1.5          & 0.03         & Pm               & AIRSS                                               & P dumbbells            \\
    Li$_2$P          & 2            & 0.02         & P2$_1$/c         & AIRSS                                               & P dumbbells            \\
    Li$_3$P $^\star$     & 3            & ~             & P6$_3$/mmc       & Known Li-P phase  \cite{Dong2007}                   & Isolated P ions       \\
    Li$_4$P          & 4            & 0.044         & C2/m             & AIRSS                                               & Isolated P ions       \\
    Li$_5$P          & 5            & 0.046         & Cmma             & AIRSS                                               & Isolated P ions       \\
    Li$_6$P          & 6            & 0.033         & Ccce             & AIRSS                                               & Isolated P ions       \\
    Li $^\star$          & ~            & ~             & Im$\bar3$m       & ~                                                   & ~                     \\ \hline
    \end{tabular}}
\end{table*}

For $0 \leq x \leq 0.5$ in Li$_x$P structures are composed mainly by tubes, cages and 3-D networks where threefold P bonding is mainly favoured. 
The least lithiated phase found on the hull is LiP$_7$ which shows tubular helices of connected P$_7$ cages  along the [001] axis. LiP$_6$--R$\bar3$m  and LiP$_5$--Pna2$_1$ exhibit relatively similar structures formed by a 3-D networks with the majority of the P ions threefold bonded.
The  next structure found on the convex hull is Li$_3$P$_7$, where the phosphorus tubes are broken, forming isolated P$_7$ cages dispersed in the 3D structure. 

In the $0.5 < x < 1.33$ region the structures are significantly different, tending to form chains and broken chains. 
The structure of Li$_x$P$_x$, x=5-9,  has recently received attention in the context of inorganic double-helix structures \cite{Ivanov2012}, where it was shown that AIRSS predicts the P2$_1$/c symmetric Li$_1$P$_1$ bulk phase; moreover, the stability of an isolated double-helix was demonstrated by phonon calculations. 
Two phases are found very close to the hull in this region, Li$_5$P$_4$ and  Li$_4$P$_3$.   Li$_4$P$_3$--P2$_1$2$_1$2$_1$ is an AIRSS structure with a formation enthalpy 4 meV/f.u. above the tie-line, a difference which is within DFT accuracy. 
Li$_5$P$_4$--C2/m  was obtained by swapping ions from Na$_5$As$_4$ \cite{Ozisik2011} and it is found  10 meV/f.u. above the convex hull. 
Both structures are formed by three (Li$_4$P$_3$) and four-bonded (Li$_5$P$_4$) in-plane chains, see Figure \ref{fig1} for an illustration. We have explored the possible mechanical stability of Li$_4$P$_3$--P2$_1$2$_1$2$_1$  by performing a phonon calculation, the calculated phonon dispersion is presented in Figure \ref{Li4P3phonon}.  
\begin{figure}
 \begin{center}
  \includegraphics[width=0.9\linewidth]{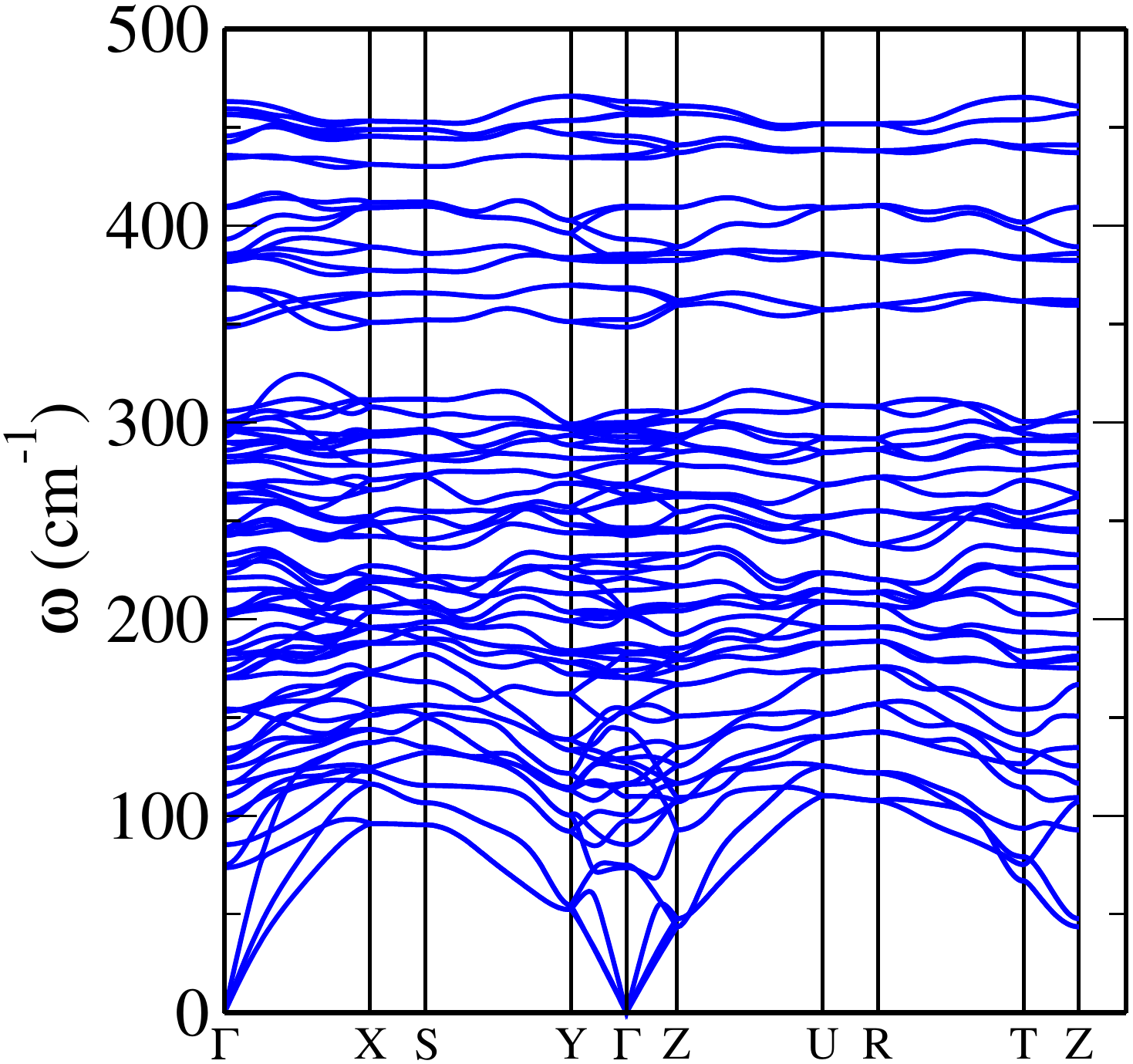}
  \caption[Phonon dispersion curve of Li4P3]{Phonon dispersion curve of  Li$_4$P$_3$-- P2$_1$2$_1$2$_1$. The absence of any imaginary frequency in the Brillouin zone confirms      the stability of a structure in terms of lattice dynamics.}
  \label{Li4P3phonon}
 \end{center}
\end{figure}
The stability of a structure in terms of lattice dynamics is confirmed by the absence of any imaginary frequency in the Brillouin zone.

For $1.33 < x \leq 2$ two structures are found by AIRSS,  Li$_2$P--P2$_1$/c and Li$_3$P$_4$--C2/m, which form P-P dumbbells. 
Dumbbell formation in Li-X (X=S,Si,Ge) systems, has played an important role in the interpretation of the electrochemical behaviour in terms of structure transformation \cite{Hyeyoung20015,Morris2014,Jung2015(MorrisGe)}.  

For concentrations larger that x=2 the P structures are formed by isolated P ions.  The most lithiated phase found on the convex hull is Li$_3$P  \cite{Brauer1937}, a phase which is generally observed at the end of discharge in electrochemical experiments \cite{Park2007,Qian2012,Qian2013}.
Three more structures are found by the AIRSS searches for $x>3$, Li$_4$P, Li$_5$P and Li$_6$P, all of them composed of isolated P atoms.

\begin{figure}
\begin{center}
\includegraphics[width=0.9\linewidth]{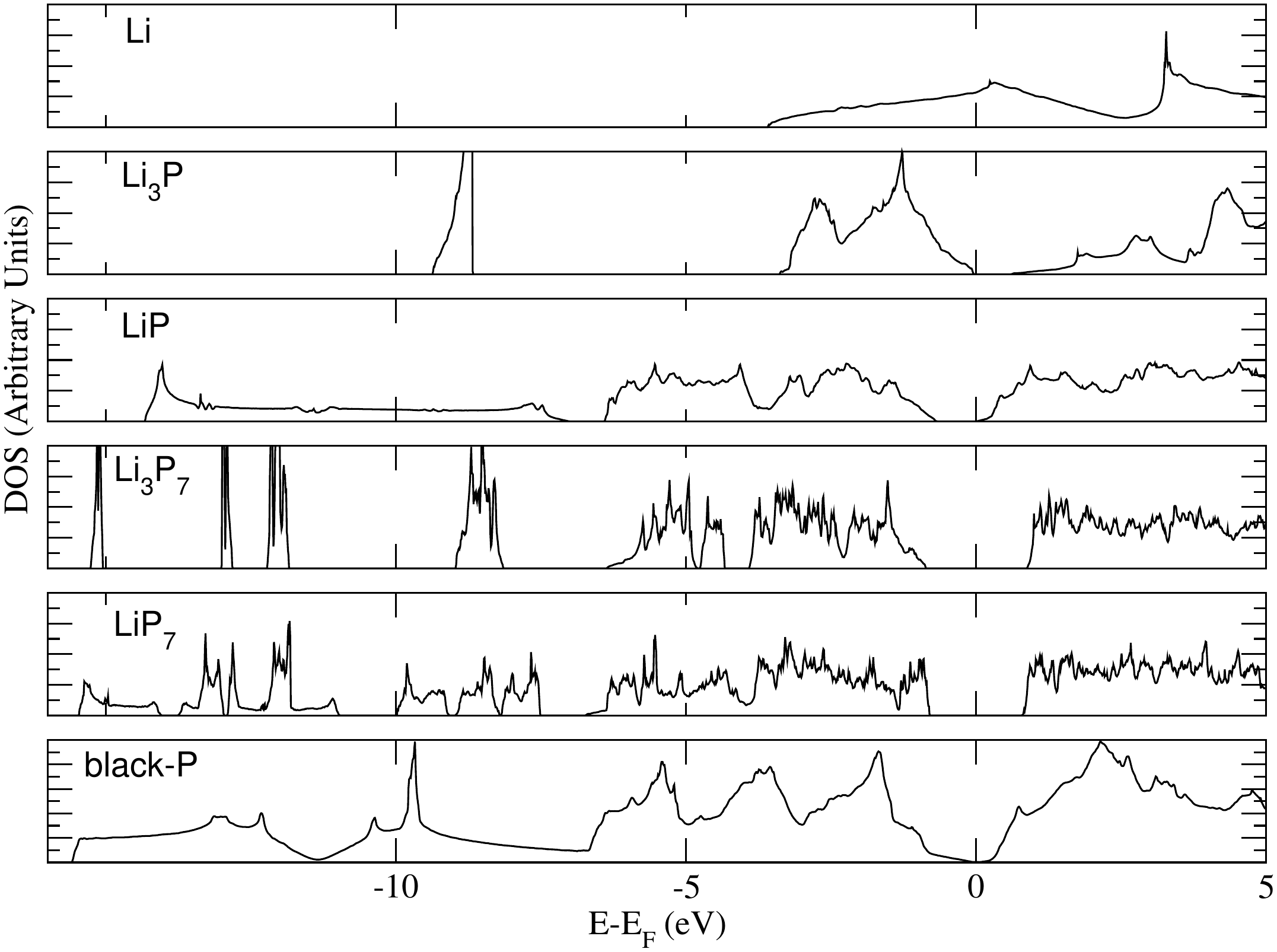}
\caption[eDOS of Li-P phases found on the convex hull.]{Total electronic density of states of the Li-P phases found on the convex hull. The Li-P structures exhibit a semiconductor-like eDOS even for high Li concentrations.}
\label{LiPDOS}
\end{center}
\end{figure}

The electronic density of states (eDOS) of the structures found on the convex hull were calculated with the OptaDOS code \cite{Morris2014b} and are shown in Figure \ref{LiPDOS}. 
All structures, except for Li, show a semiconducting-like eDOS, which is surprising especially for  the phases with high lithium concentration. 
 
\begin{figure}
\begin{center}
\includegraphics[width=\linewidth]{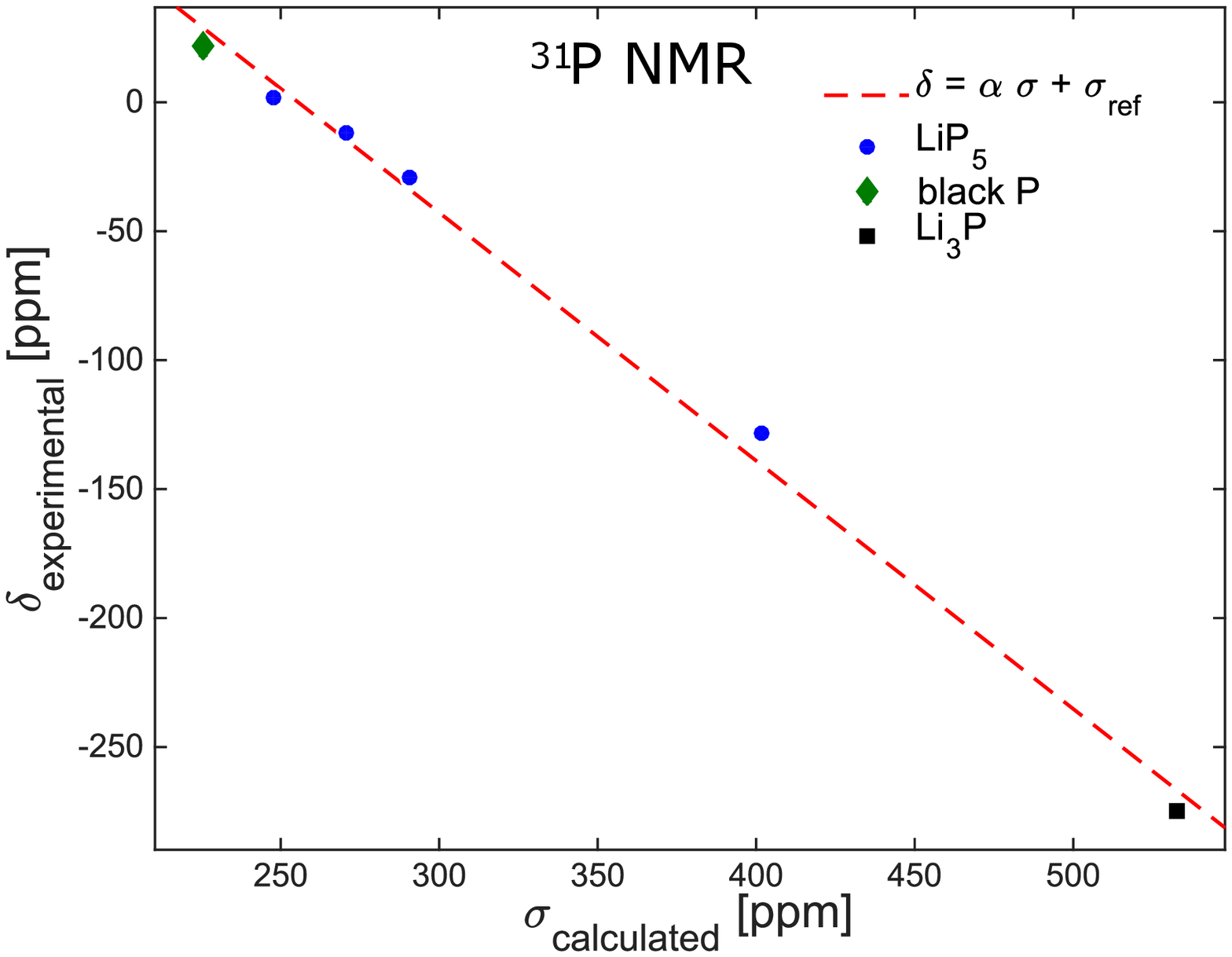}
\caption[]{Correlation between the $^{31}$P NMR calculated chemical shielding, $\sigma_{calc.}$, and experimental chemical shifts, $\delta_{exp.}$,  referenced relative to an 85\% H$_3$PO$_4$ aqueous solution for LiP$_5$\cite{Gunne1999} Li$_3$P \cite{marino2012} and black P \cite{ramireddy2015} . The data was fitted to a linear function $\delta_{exp.}=\alpha\sigma_{calc.}+\sigma_{ref.}$ with resultant fitting parameters $\alpha=-0.96\pm0.1$  and $\sigma_{ref.}=245.9\pm34.1$. The deviation of $\alpha$ from the ideal value of $-1$ is well known when correlating between calculated shielding and experimental shifts (See  Ref. \cite {laskowski2013} for details), the obtained $\sigma_{ref.}$ was used  to reference the presented NMR results (See Figures \ref{LiPnmr} and \ref{NaPnmr}).}  
\label{LiPnmrCorrelation}
\end{center}
\end{figure}

The experimental ability of measuring NMR spectra during charge and discharge, can be an extremely powerful tool to elucidate the structural evolution of the anode during the lithiation \cite{Ogata2014}.
We have calculated the phosphorus chemical shielding for the stable structures of the Li-P system. 
We have included  the LiP$_5$ Pna2$_1$  chemical shielding calculation for comparison  with the experimental data reported in Ref. \citenum{Gunne1999}. 
A plot of the correlation between the calculated  and experimental NMR parameters of LiP$_5$\cite{Gunne1999}, Li$_3$P\cite{marino2012} and black P\cite{ramireddy2015} is presented in Figure \ref{LiPnmrCorrelation}, where a good correlation is seen between experimental and calculated values.

\begin{figure*}
\begin{center}
\includegraphics[width=0.65\linewidth]{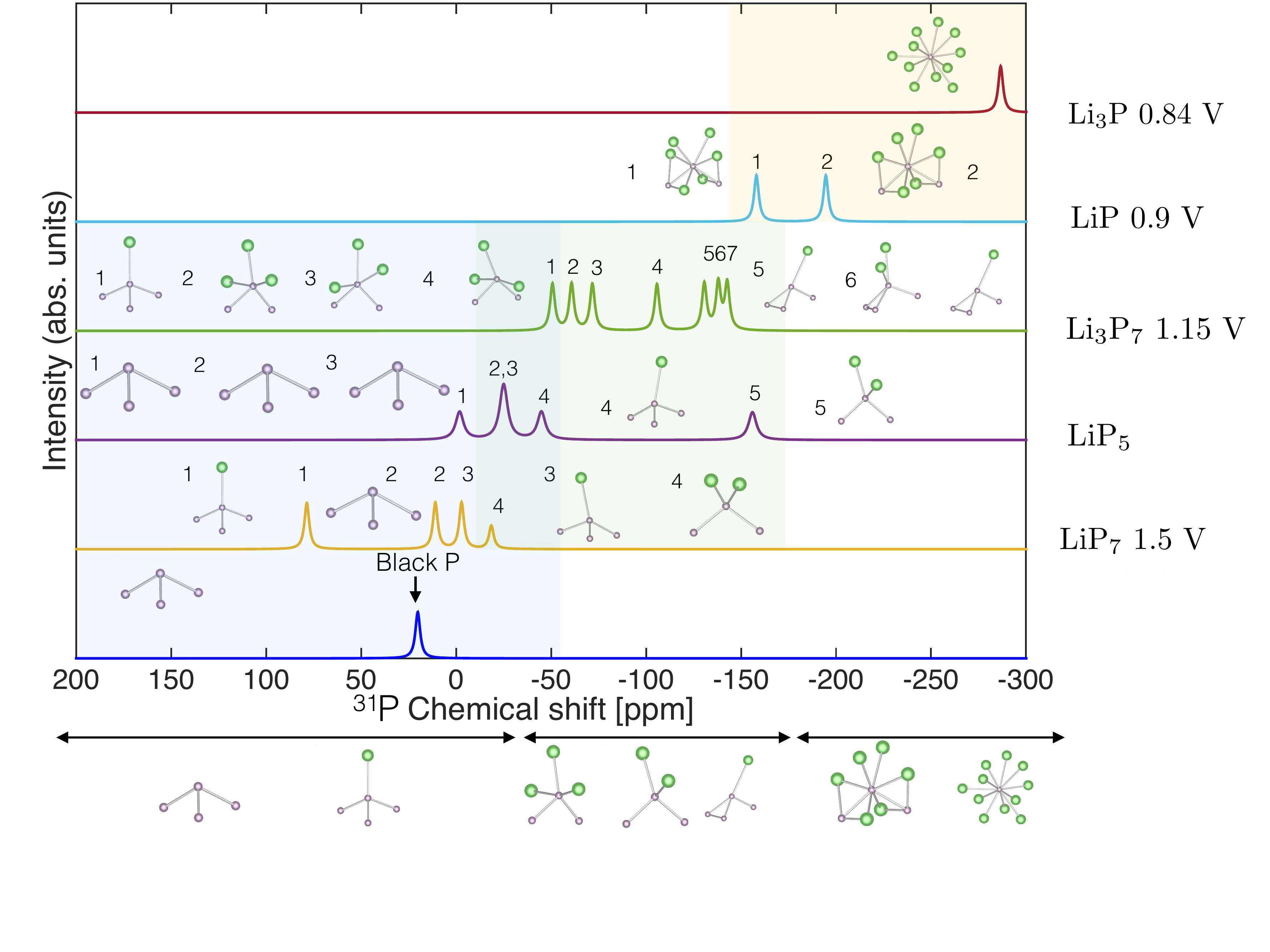}
\caption[]{Calculated $^{31}$P NMR chemical shifts, referenced using $\sigma_{\mathrm ref.}$ obtained from Figure \ref{LiPnmrCorrelation}, for various  Li-P compounds showing the change in chemical shielding as the local environment of
phosphorus changes. For visualisation purposes a Lorentzian broadening is assigned to the calculated $^{31}$P NMR parameters. For each crystallographic site a cluster with a radius of 3 \textrm{\AA} is shown and labeled accordingly. We have coloured the background to guide the eye between the three regions, above -45 ppm, below -155 ppm and an intermediate region. Above -45 ppm ppm predominantly the structures have three phosphorus nearest neighbours (NNs)  and one or no lithium NN. Below -155  ppm the phosphorus has more than six lithium NNs. In the intermediate region the phosphorus ions tends to have four or five NNs of which two to three are
phosphorus atoms.}
\label{LiPnmr}
\end{center}
\end{figure*}

The resulting NMR parameters of all the structures are illustrated in Figure \ref{LiPnmr}.
A general trend is observed in chemical shielding, where the latter increases with the Li concentration in Li$_x$P.  
We identify three main regions in the chemical shielding described in Figure \ref{LiPnmr} which can be roughly related to the amount of lithium and phosphorus nearest neighbours (See caption in Figure \ref{LiPnmr} for detailed description).

\subsection{Sodium phosphide }

\begin{figure*}
\begin{center}
\includegraphics[width=0.65\linewidth]{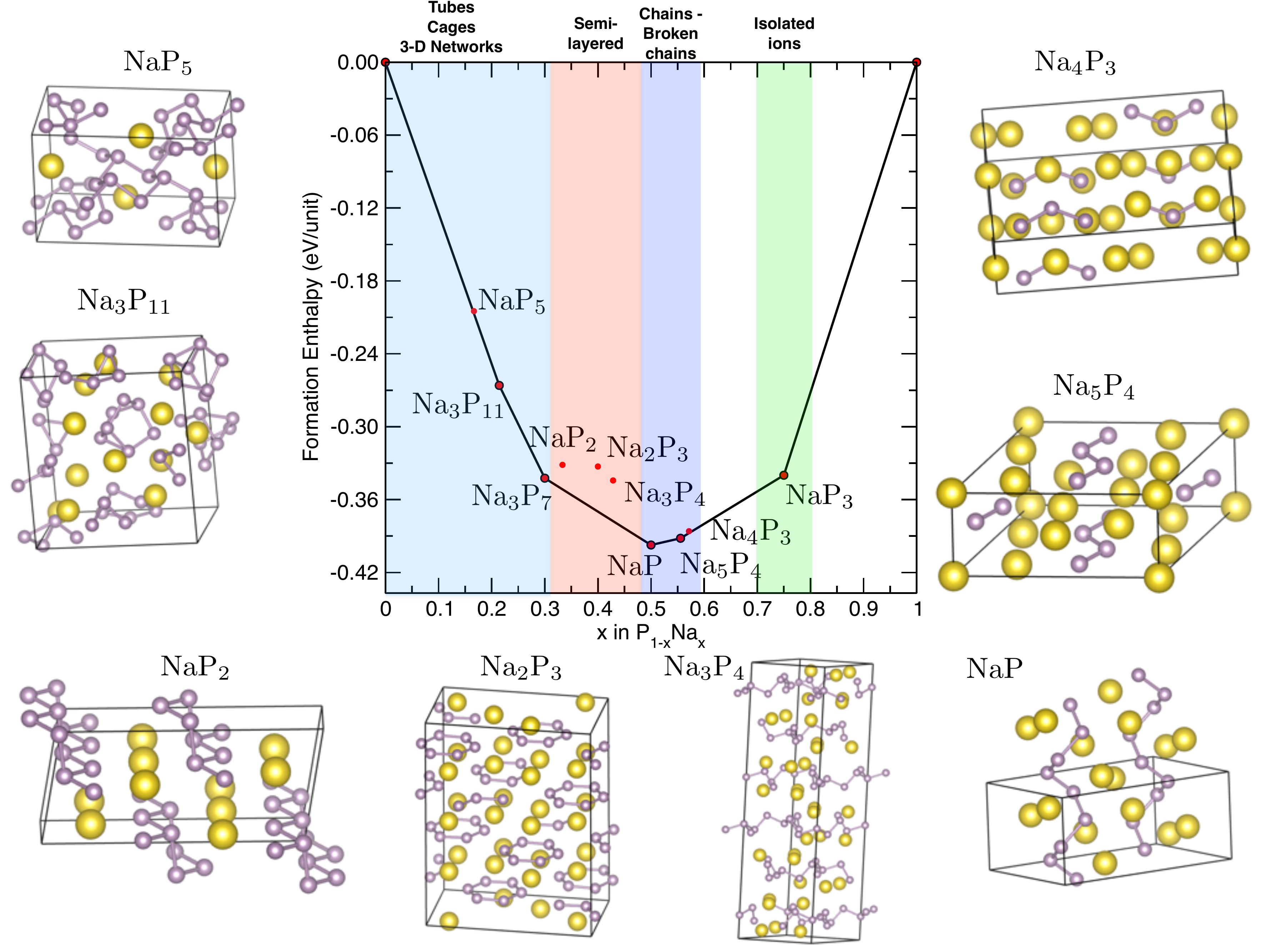}
\caption[Convex hull of Na-P system.]{Formation enthalpy per atom vs the fractional sodium concentration in the Na-P compound. The convex hull (tie-line) is constructed by joining the stable structures obtained by the searches.}
\label{NaPhull}
\end{center}
\end{figure*}

\begin{figure}
\begin{center}
\includegraphics[width=0.45\textwidth]{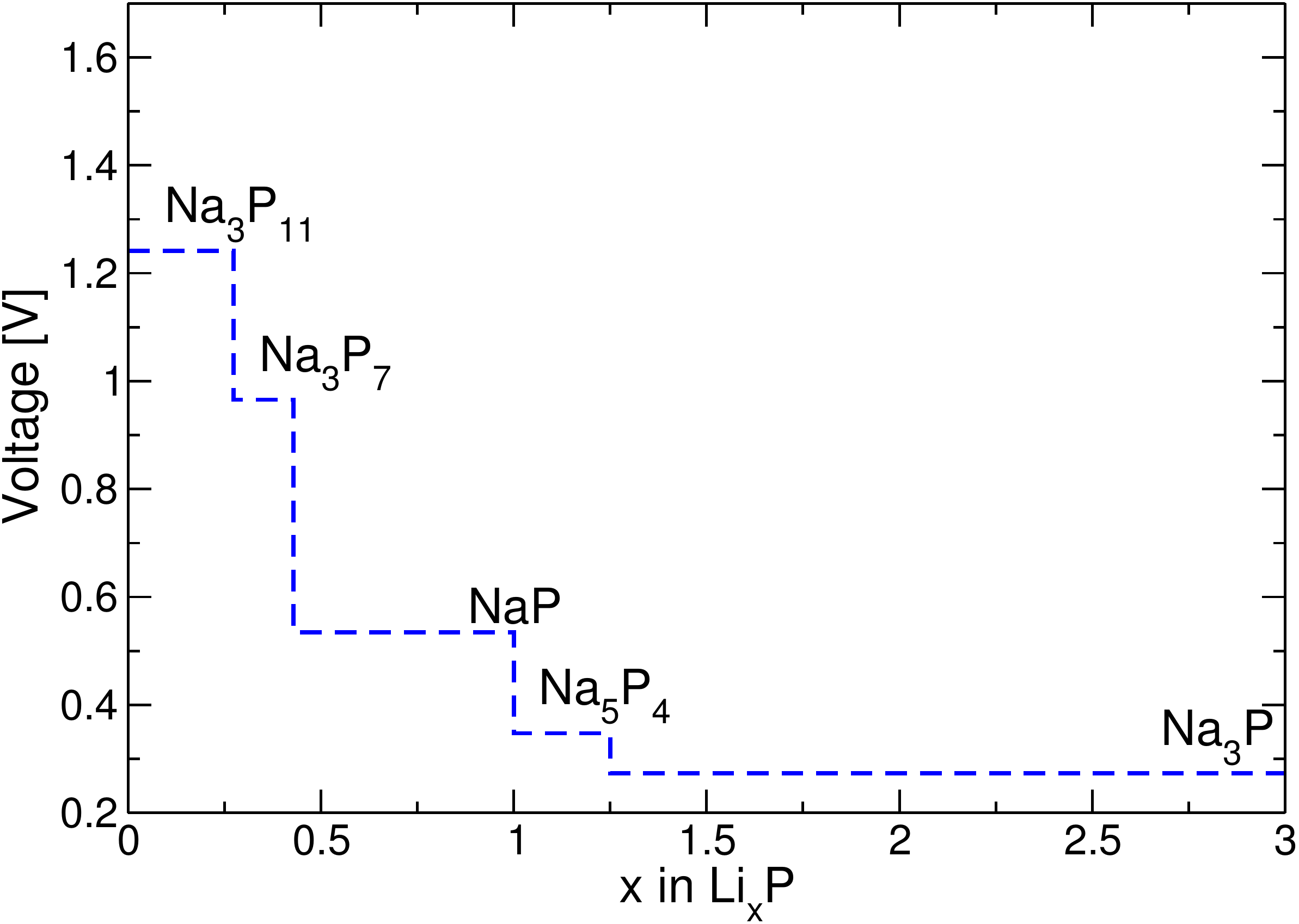}
\caption[Voltage profile of Na-P system.]{Average voltages  relative to sodium calculated for the structures found on the convex hull (Figure \ref{NaPhull}). }
\label{NaPvolts}
\end{center}
\end{figure}

Na-P forms similar structures to those found for Li-P, as expected due to their similar chemistry. However, the convex hull of the NaP system, as shown in Figure \ref{NaPhull}, exhibits two main differences: first, the Li$_1$P$_1$ phase has a lower formation energy than Na$_1$P$_1$ by approximately -0.125 eV, the second is that the Li$_3$P phase has  lower formation energy than  Li$_1$P$_1$ by -0.125 eV, whereas  Na$_3$P has higher formation energy than Na$_1$P$_1$ by 0.05 eV/f.u.. These differences are manifested in the calculated average voltages (see Figures \ref{NaPvolts} and \ref{fig3}), where the Na-P voltage profile drops to lower values at high Na concentrations.
The stable phases predicted by the DFT calculations are summarised in Table \ref{NaPtable}. 

\begin{table*} 
\centering
\caption[Na-P  phases found on the convex hull.]{Description of the experimental and predicted Na$_x$P phases. We indicate with a star ($\star$) the stable phases which are found on the convex hull. The Na-P structures show similar ion arrangements as those observed in Li-P (see Figure \ref{fig1} for illustration), with differences in the $0.45 < x < 1$ region and the absence of P dumbbells .     } \label{NaPtable}
\resizebox{\linewidth}{!}{
  \begin{tabular}{lccclc}
    \hline
    Stoichometry        & x in Li$_x$P & Distance from & Space            & Structure origin                              & Description                  \\
    ~                   & ~            & the hull [eV/f.u.] & group            & ~                                             & ~                            \\ \hline
    black-P  $^\star$             & 0            & ~             & Cmca             & ~                                             & ~                            \\
    NaP$_5$             & 0.2          & 0.002         & Pnma             & Swapping from LiP$_5$ \cite{Schnering1972}    & P 3-D network                \\
    Na$_3$P$_{11}$ $^\star$ & 0.273        & ~             & Pbcn             & Known Na-P phase \cite{Wichelhaus1973}        & P$_{11}$ cages               \\
    Na$_3$P$_7$ $^\star$    & 0.429        & ~             & P2$_1$2$_1$2$_1$ & Known Na-P phase \cite{Honle1983}             & P$_{7}$ cages                \\
    NaP$_2$             & 0.5          & 0.02          & C2/m             & Swapping from KSb$_2$ \cite{Rehr1995}         & Black P - like broken layers \\
    Na$_2$P$_3$         & 0.667        & 0.037         & Fddd             & Swapping from K$_4$P$_6$ \cite{Abicht1984}    & P six-fold rings             \\
    Na$_3$P$_4$         & 0.75         & 0.034         & R$\bar3$c        & AIRSS                                         & In-plane connected chains    \\
    NaP $^\star$            & 1            & ~             & P2$_1$/c         & Known Na-P phase \cite{Honle1981}             & P helix                      \\
    Na$_5$P$_4$ $^\star$    & 1.25         & ~             & C2/m             & Swapping from Na$_5$As$_4$ \cite{Ozisik2011}  & 4 P zig-zag chains           \\
    Na$_4$P$_3$         & 1.333        & 0.002         & Cmcm             & Swapping from K$_4$P$_3$ \cite{Schnering1989} & 4 P zig-zag chains           \\
    Na$_3$P $^\star$        & 3            & ~             & P6$_3$cm         & Swapping from Na$_3$As \cite{Ozisik2011}      & Isolated P ions              \\
    Na$_3$P             & 3            & 0.005         & P6$_3$/mmc       & Known Na-P \cite{Brauer1937}                  & Isolated P ions              \\
    Na $^\star$             & ~            & ~             & Im$\bar3$m            & ~                                             & ~                            \\ \hline
    \end{tabular}}
\end{table*}

The average voltages were calculated for Na-P and are shown in Figure \ref{NaPvolts}. 

The least sodiated Na-P structure found in the Na-P convex hull construction is a locally stable NaP$_5$-- Pnma phase, which was obtained by swapping species from LiP$_5$ \cite{Schnering1972}. Increasing in sodium content, two known phases are found on the convex hull,  Na$_3$P$_{11}$--Pbcn \cite{Wichelhaus1973} and Na$_3$P$_7$--P2$_1$2$_1$2$_1$ \cite{Honle1983}. In the $0.45 < x < 1$ region we find  three structures with rather different ionic arrangements, exhibiting broken black P - like layers (NaP$_2$-- C2/m \cite{Rehr1995}),  P six-fold rings (Na$_2$P$_3$--Fddd  \cite{Abicht1984}) and in-plane connected chains ( Na$_3$P$_4$--R$\bar3$c predicted by AIRSS). For $x>1$ the structures show similar arrangements as in the Li-P system, although, unlike in Li-P, Na-P does not seem to favour dumbbell formations. The Na$_5$P$_4$--C2/m obtained by swapping atoms from  Na$_5$As$_4$ \cite{Ozisik2011} exhibits a layered structure consisting of Na sheets separated by four-bounded  in-plane P chains. This new phase is predicted to be thermodynamically stable by our calculations. Furthermore, its calculated phonon dispersion curve confirms the stability of the phase in terms of lattice dynamics. 
\begin{figure}
\begin{center}
\includegraphics[width=0.9\linewidth]{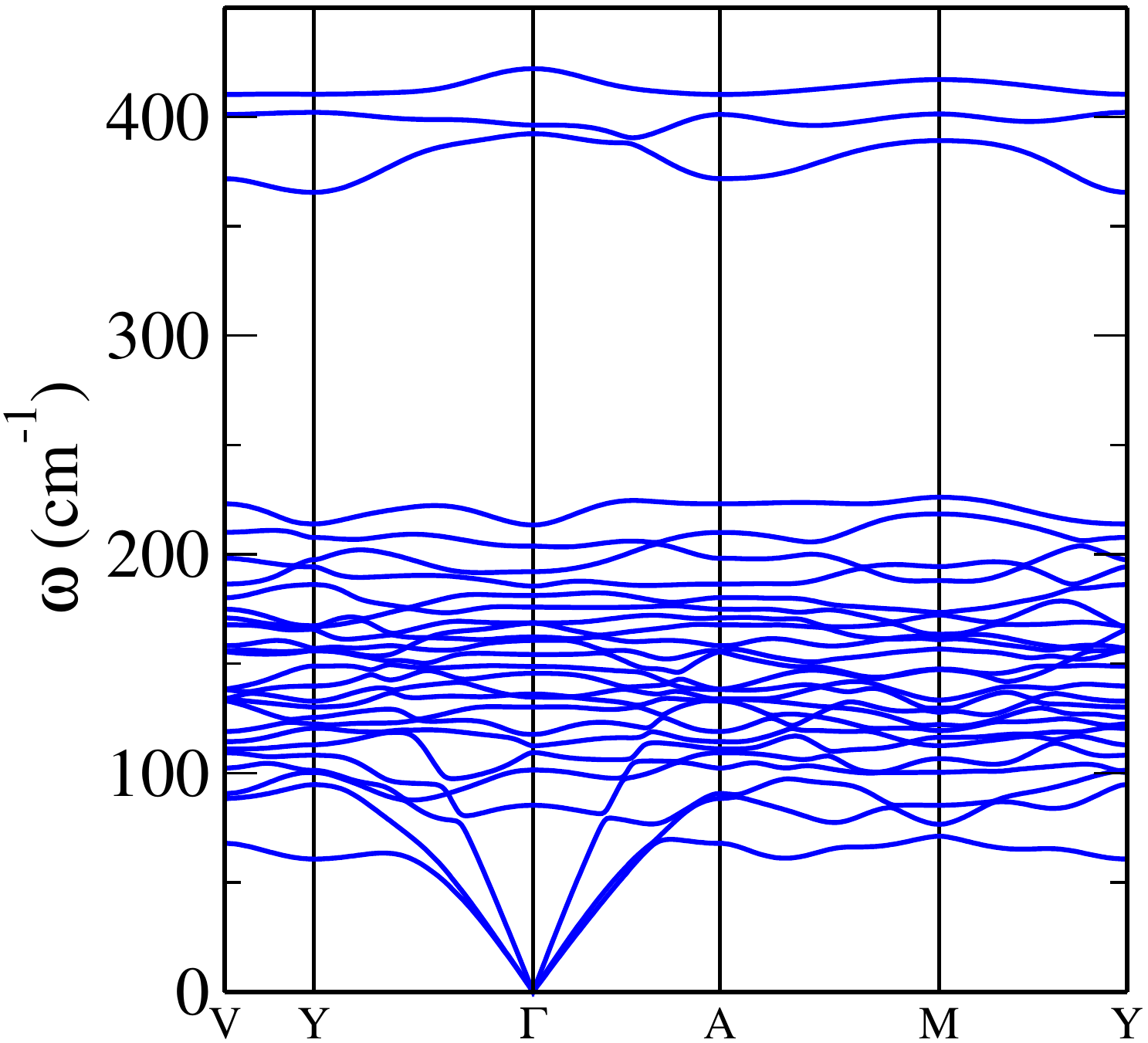}
\caption[Phonon dispersion curve of Na5P4]{Phonon dispersion curve of  Na$_5$P$_4$--C2/m. The absence of any imaginary frequency in the Brillouin zone confirms the stability of a structure in terms of lattice dynamics.}
\label{Li4P3phonon}
\end{center}
\end{figure} 

As in the Li-P system,  the Na-P phases exhibit a semiconducting behaviour, except for the  Na$_5$P$_4$ phase which shows a finite value of eDOS at the Fermi energy. 

\begin{figure}
\begin{center}
\includegraphics[width=0.9\linewidth]{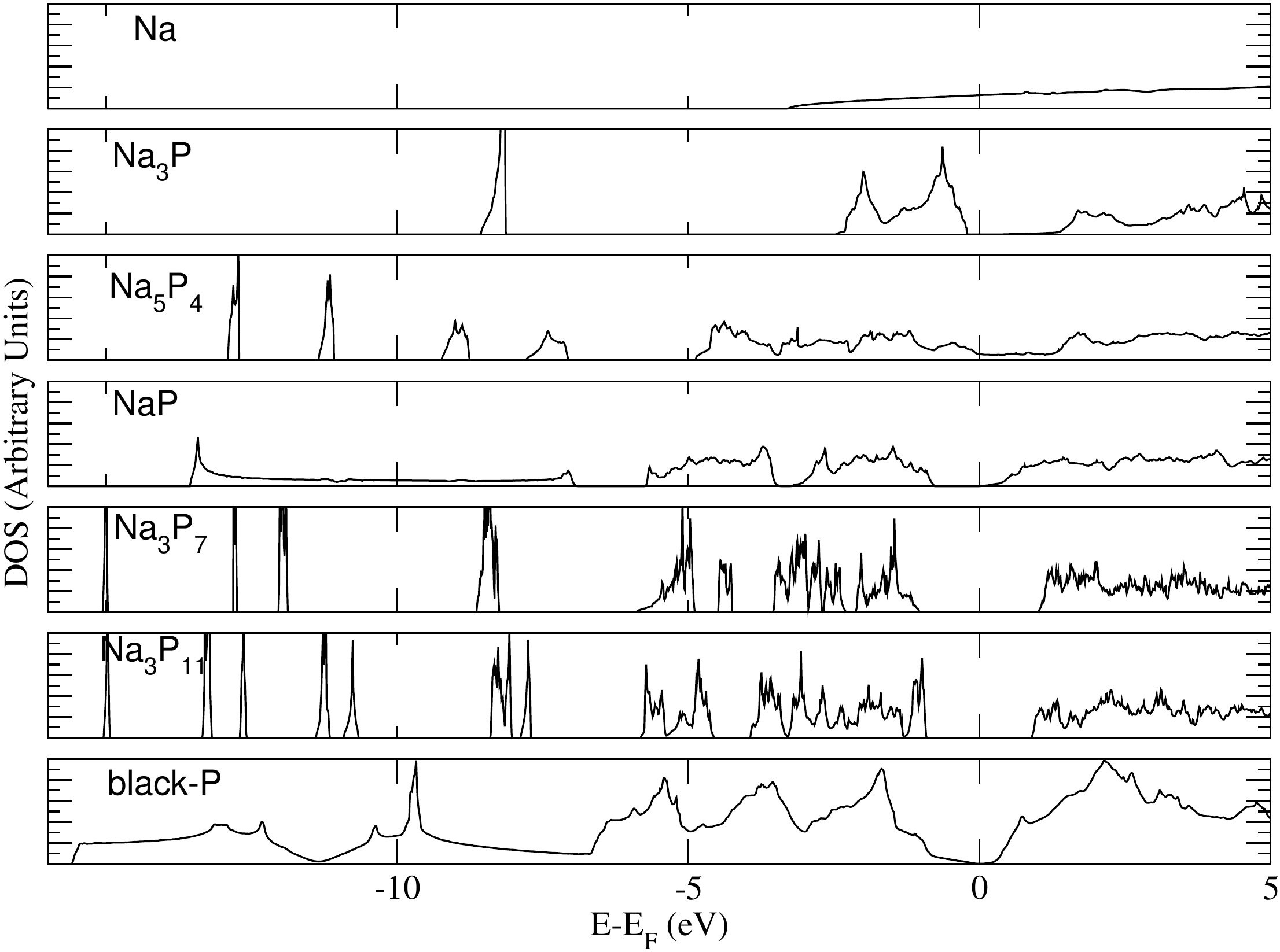}
\caption[eDOS of Na-P phases found on the convex hull.]{Total electronic density of states of the Na-P phases found on the convex hull. The Na-P phases exhibit a semiconductor-like eDOS, except  for the Na$_5$P$_4$ which has a finite value of eDOS at the Fermi level. }
\label{NaPDOS}
\end{center}
\end{figure}

\begin{figure*}
\begin{center}
\includegraphics[width=0.65\linewidth]{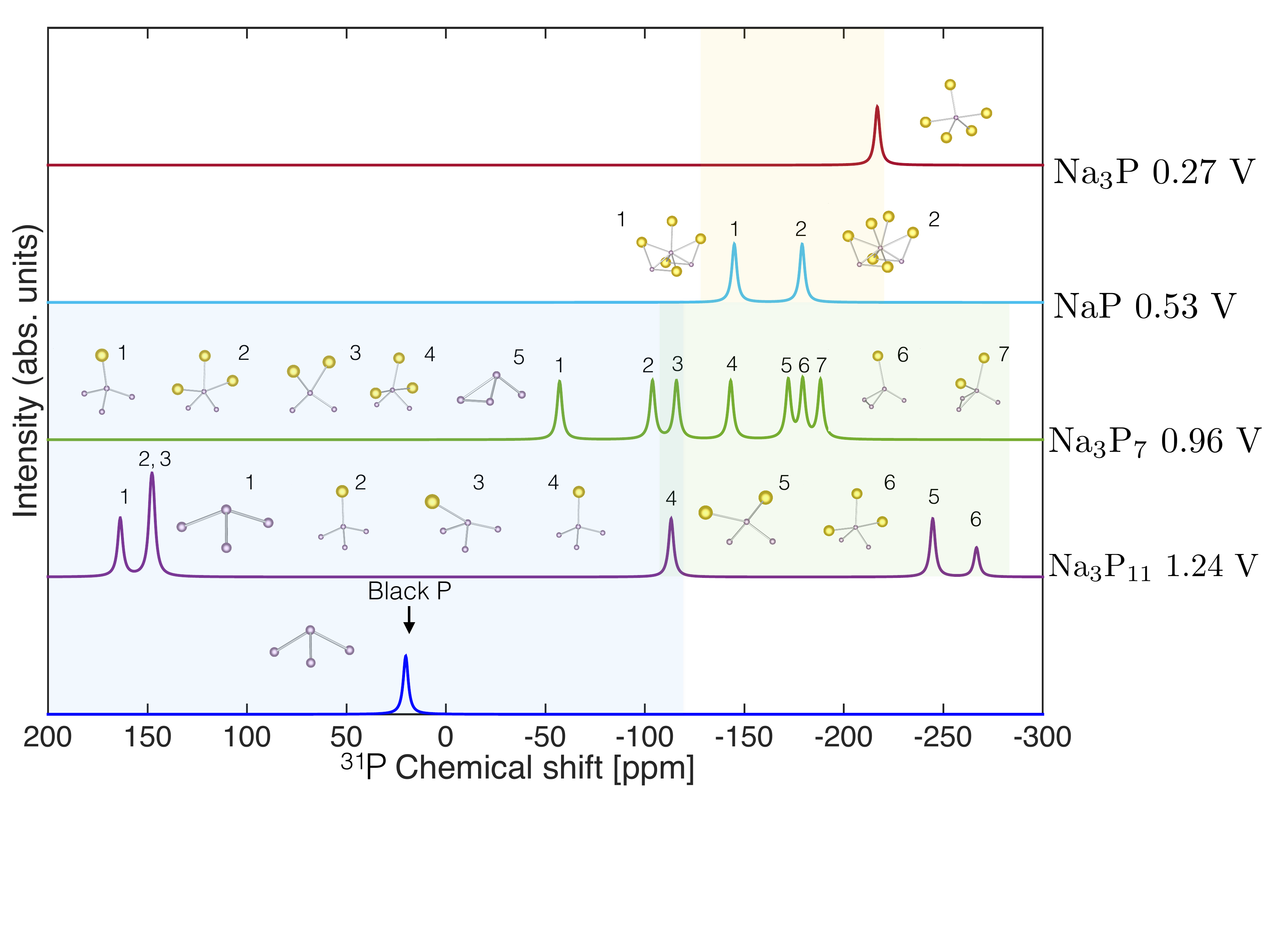}
\caption[]{Calculated $^{31}$P NMR chemical shifts for various  Na-P compounds showing the change in chemical shielding as the local environment of
phosphorus changes. For visualisation purposes a Lorentzian broadening is assigned to the calculated $^{31}$P NMR parameters. For each crystallographic site a cluster with a radius of 3 \textrm{\AA} is shown and labeled accordingly. The background has been coloured as in Figure \ref{LiPnmr} to emphasise regions in the chemical shift associated with specific atomic arrangements. Despite the similarities to the Li-P,  it may be more difficult to experimentally differentiate the
mid and high sodiated regions due to a more similar chemical shielding.}
\label{NaPnmr}
\end{center}
\end{figure*}

\subsection{Aluminium doping of phosphorus} \label{Al doping}

In order to suggest a way for improving the electrical conductivity of phosphorus anodes we have tested the effect of different extrinsic dopants on the electronic DOS of Li-P compounds by performing interstitial defects AIRSS searches. The initial generated structures were composed of the underlying perfect crystal plus the interstitial element positioned randomly. Consequently,  the ionic positions were relaxed keeping the lattice vectors fixed. The electronic DOS of the lowest-energy structures were then calculated using OptaDOS. 

\begin{figure}
        \centering
        \begin{subfigure}{0.9\linewidth}
                \centering
                \includegraphics[width=\textwidth]{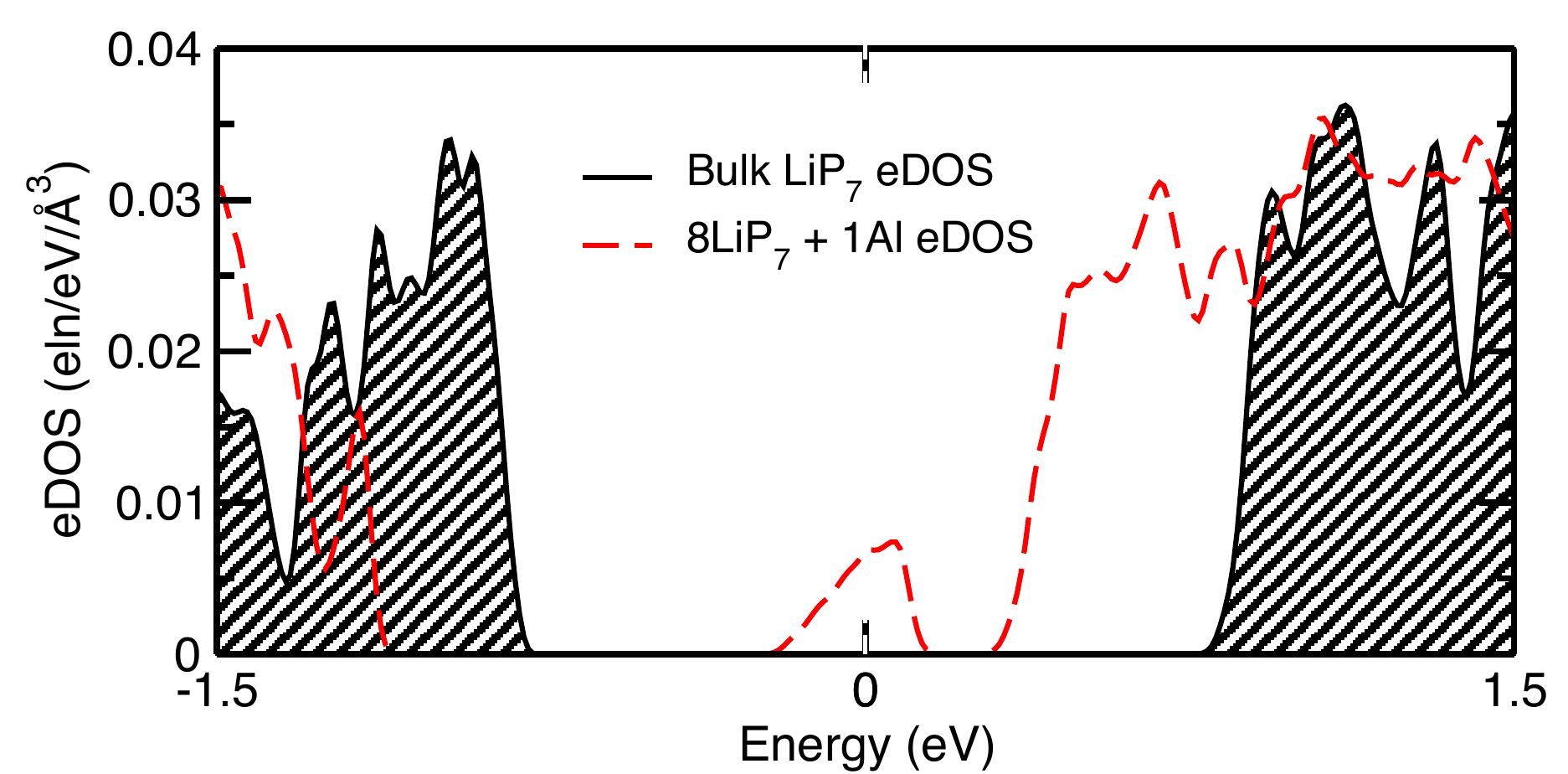}
                \caption{8LiP$_7$ + 1Al eDOS}
        \end{subfigure}\\
        \begin{subfigure}{0.9\linewidth}
                \centering
                \includegraphics[width=\textwidth]{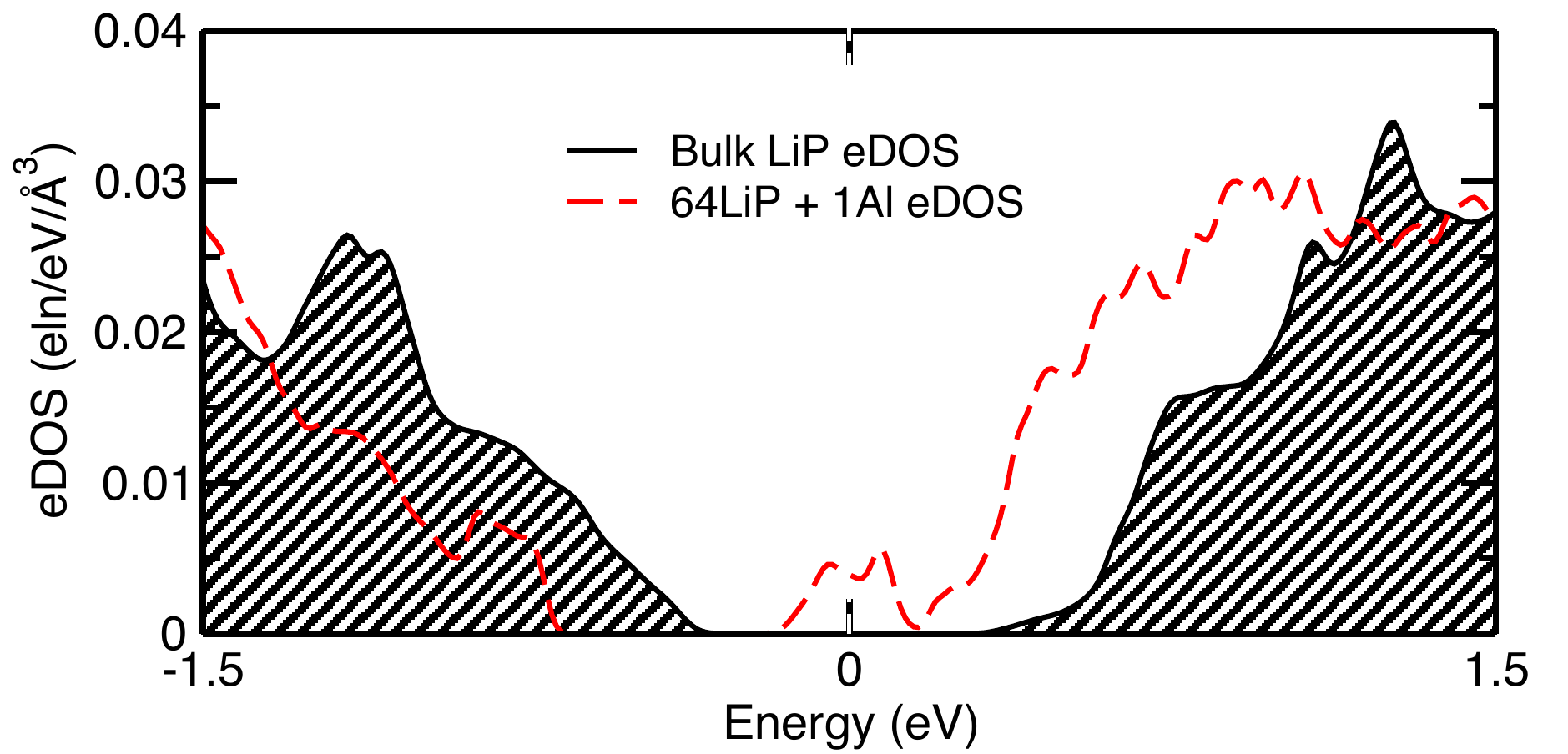}
                \caption{64LiP + 1Al eDOS}
        \end{subfigure}\\
        \begin{subfigure}{0.9\linewidth}
                \centering
                \includegraphics[width=\textwidth]{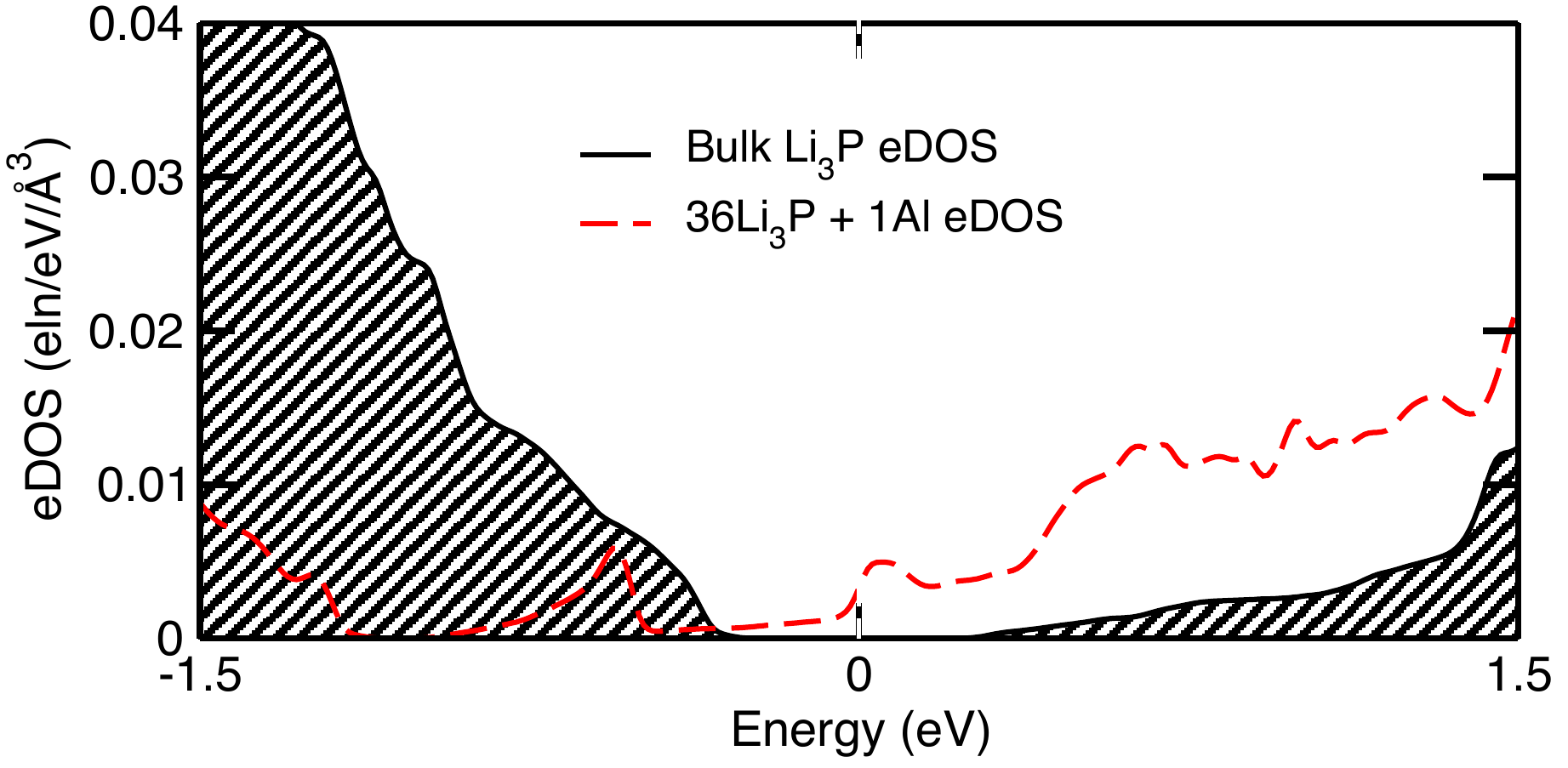}
                \caption{36Li$_3$P + 1Al eDOS}
        \end{subfigure}  
\caption{Electronic density of states in the vicinity of E$_F$ for different Li-P compounds found on the convex hull (black line - dashed background)  and Li-P with one aluminium interstitial defect (dashed red line). A finite electronic DOS is found at the Fermi energy for the Li-P compounds + Al, contrary to pristine Li-P compounds which exhibit band gaps around the Fermi level.  }
\label{eDOSLiPAL}        
\end{figure}

Silicon and aluminium interstitial defect searches were carried out in a $2 \times 2 \times 1$  LiP supercell composed of 32 LiP formula units, we denote these structures as 32LiP + 1Si and 32LiP + 1Al, respectively. The eDOS  calculation revealed that the aluminium point defect introduces electronic states at the Fermi energy (E$_F$), whereas the silicon defect introduces states within the band gap but with the eDOS remaining zero at E$_F$. To further investigate the effect of Al doping, AIRSS searches were performed in larger Li-P cells with different Li concentrations. The cells were chosen to be large enough to allow a maximum stress of ca. 0.5 GPa.  Figure \ref{eDOSLiPAL} shows the resulting eDOS of  8LiP$_7$ + 1Al, 64LiP + 1Al and 36Li$_3$P + 1Al for the lowest-energy structure resulting from the searches.    
From Figure \ref{eDOSLiPAL}, we learn that the formed Li-P compounds with different Li concentrations exhibit finite electronic DOS at E$_F$, suggesting that doping phosphorus with aluminium could increase the electronic conductivity of the anode, thus improving its performance.

\section{Discussion}

We have presented a study of Li-P and Na-P systems using AIRSS and atomic species swapping of ICSD structures. We have shown that the combination of the two methods allows us to have access not only to the ground state structures, but also metastable phases found close to the convex hull. These structures might form at room temperature and non-equilibrium conditions, \eg, during lithiation/sodiation. The aim of this work is to elucidate the structural evolution of  phosphorus anodes during the lithiation/sodiation  as well as to give insights into their electronic structure, some of these aspects are discussed below.

The method of AIRSS + atomic species swapping has been shown to predict a variety of locally stable phases in the Li-P system (see Table \ref{LiPtable} for full description). Combining the known phases with those predicted in this work, we are enabled to catalogue  phosphorus ionic arrangements according to their lithium concentration. This has proved to be extremely valuable when attempting to understand electrochemical processes, as has been recently shown in Ref. \citenum{Hyeyoung20015} for LiS batteries. Our findings suggest that the lithiation mechanism proposed in  Ref. \citenum{Park2007}, Black P $\rightarrow$ Li$_x$P$\rightarrow$LiP$\rightarrow$Li$_2$P$\rightarrow$Li$_3$P, could be reinterpreted in terms of tubes, cages and 3-D networks $\rightarrow$ chains and broken chains $\rightarrow$ P dumbbells $\rightarrow$ isolated P ions. Moreover, phases found by our structure searching can clarify possible intermediate structures in a more robust way. Park \etal \cite{Park2007} predicted the existence of a metastable Li$_2$P structure based on the appearance of a XRD peak at $2\theta \approx 22.5^\circ$  which corresponded to a molar ratio of Li:P 2 at 0.63 V. Our Li$_2$P--P2$_1$/c structure exhibits a high intensity predominant peak at $2\theta \approx 25^\circ$, a discrepancy which can be attributed to the difference between the DFT and experimental lattice parameters.  

The convex hull of the Na-P system predicts a locally stable NaP$_5$--Pnma phase which is very close to the convex hull; this phase has been synthesised at high-pressure \cite{Chen2004}. A  new phase,  Na$_5$P$_4$, with C2/m symmetry  is predicted to be stable by the convex hull construction. The phonon dispersion of the stable phase, Na$_5$P$_4$--C2/m, and the  Li$_4$P$_3$--P2$_1$2$_1$2$_1$ metastable phase found very close to the convex hull suggest that these predicted structures are mechanically stable and might be observed in future experiments. 

NMR chemical shielding calculations reveal a general trend in the local environment change of both Li-P and Na-P systems as lithium/sodium content is increased. For Li-P the chemical shielding range was roughly divided in three regions, where each region was correlated to distinct local ionic arrangements. These calculations were driven by the experimental ability of measuring NMR shifts, where the assignment of the local environments of the probed ion can be particularly challenging.  Na-P shows a similar trend, but due to overlap between the regions it may be more difficult to assign experimental data.

Li-P and Na-P exhibit relatively high average voltage profiles, which in principle leads to a lower voltage of the full cell and a reduced energy density. The Na-P voltage profile differs from the Li-P profile, the voltage drops to 0.28 V in the case of the Na$_3$P phase, whereas in the case of Li-P it drops to 0.8 V at the same lithium concentration. Despite this disadvantage, high voltages prevent the formation of Li dendrites, thus enhancing the safety of the battery. A second advantage of high voltages versus lithium metal, is the prevention of electrochemical reduction of the electrolyte as SEI forms, which can improve the cyclability of the battery \cite{Nitta2014}. 

Despite several advantages, pure phosphorus shows a relatively poor cyclability  \cite{Park2007,Sun2012}. Park \etal \cite{Park2007} attributed the low performance of phosphorus anodes to its low electronic conductivity.  Sun \etal \cite{Sun2012} showed that black P samples exhibit  good conductivity properties, and put the low performance of the anode down to the non-crystallinity of the samples.  Our results show that even for low concentrations of Li, Li-P compounds can exhibit a relatively large band gap, e.g. 1.7 eV for LiP$_7$, compared to the experimental 0.33 eV of black P, hinting than the conducting properties of black P can be worsened as the anode is lithiated. In order to address this, we have sought to reduce the band gap of Li-P compounds by doping them with aluminium. Furthermore, we have performed a preliminary study on the effect of Ge and Ga doping on Li-P compounds electronic structure, where results show a similar behaviour as Si and Al respectively. 32LiP +1Ga exhibits a larger eDOS at E$_F$ compared to 32LiP +1Al, 22.88 eln/cell  and 10.78 eln/cell respectively. However, the lighter weight and high abundance of aluminium make it a promising dopant.         

\section{Summary}

We have presented above an \abinitio study of phosphorus anodes for Li and Na-ion batteries and proposed a theoretical  lithiation/sodiation process using the structure prediction AIRSS method. Our searches reveal the existence of a variety of metastable structures which  can can appear in out-of-equilibrium processes such as charge and discharge. In particular, a Li$_4$P$_3$--P2$_1$2$_1$2$_1$ AIRSS structure found to lie very close to the convex hull and a new Na$_5$P$_4$--C2/m structure  obtained by the species swapping method  found stable at 0 K. The dynamical stability of these structures was probed by phonon calculations. Our calculations showed a high theoretical voltage vs Li metal for Li-P, which makes phosphorus a good candidate for safe anodes at high rate charges. We have calculated $^{31}$P NMR chemical shielding and related them to local structure arrangements, which combined with future $^{31}$P NMR experiments can elucidate lithiation and sodiation mechanisms. Finally we have studied the effect of dopants on the electronic structure of Li-P compounds, where we conclude that doping the anode with aluminium can improve its electrochemical behaviour.

\bibliography{references}

\providecommand*\mcitethebibliography{\thebibliography}
\csname @ifundefined\endcsname{endmcitethebibliography}
  {\let\endmcitethebibliography\endthebibliography}{}
\begin{mcitethebibliography}{59}
\providecommand*\natexlab[1]{#1}
\providecommand*\mciteSetBstSublistMode[1]{}
\providecommand*\mciteSetBstMaxWidthForm[2]{}
\providecommand*\mciteBstWouldAddEndPuncttrue
  {\def\EndOfBibitem{\unskip.}}
\providecommand*\mciteBstWouldAddEndPunctfalse
  {\let\EndOfBibitem\relax}
\providecommand*\mciteSetBstMidEndSepPunct[3]{}
\providecommand*\mciteSetBstSublistLabelBeginEnd[3]{}
\providecommand*\EndOfBibitem{}
\mciteSetBstSublistMode{f}
\mciteSetBstMaxWidthForm{subitem}{(\alph{mcitesubitemcount})}
\mciteSetBstSublistLabelBeginEnd
  {\mcitemaxwidthsubitemform\space}
  {\relax}
  {\relax}

\bibitem[Chu and Majumdar(2012)Chu, and Majumdar]{Chu2012}
Chu,~S.; Majumdar,~A. \emph{Nature} \textbf{2012}, \emph{488}, 294--303\relax
\mciteBstWouldAddEndPuncttrue
\mciteSetBstMidEndSepPunct{\mcitedefaultmidpunct}
{\mcitedefaultendpunct}{\mcitedefaultseppunct}\relax
\EndOfBibitem
\bibitem[Scrosati et~al.(2013)Scrosati, Abraham, van Schalkwijk, and
  Hassoun]{LIBbook}
Scrosati,~B.; Abraham,~K.; van Schalkwijk,~W.; Hassoun,~J. \emph{{Lithium
  Batteries: Advanced Technologies and Applications}}; The ECS Series of Texts
  and Monographs; Wiley, 2013\relax
\mciteBstWouldAddEndPuncttrue
\mciteSetBstMidEndSepPunct{\mcitedefaultmidpunct}
{\mcitedefaultendpunct}{\mcitedefaultseppunct}\relax
\EndOfBibitem
\bibitem[Nitta and Yushin(2014)Nitta, and Yushin]{Nitta2014}
Nitta,~N.; Yushin,~G. \emph{Part. Part. Syst. Char.} \textbf{2014}, \emph{31},
  317--336\relax
\mciteBstWouldAddEndPuncttrue
\mciteSetBstMidEndSepPunct{\mcitedefaultmidpunct}
{\mcitedefaultendpunct}{\mcitedefaultseppunct}\relax
\EndOfBibitem
\bibitem[Kim et~al.(2011)Kim, Kim, and Kang]{Kim2011}
Kim,~Y.; Kim,~D.; Kang,~S. \emph{Chem. Mater.} \textbf{2011}, \emph{23},
  5388--5397\relax
\mciteBstWouldAddEndPuncttrue
\mciteSetBstMidEndSepPunct{\mcitedefaultmidpunct}
{\mcitedefaultendpunct}{\mcitedefaultseppunct}\relax
\EndOfBibitem
\bibitem[McDowell et~al.(2013)McDowell, Lee, Nix, and Cui]{McDowell2013}
McDowell,~M.~T.; Lee,~S.~W.; Nix,~W.~D.; Cui,~Y. \emph{Adv. Mater.}
  \textbf{2013}, \emph{25}, 4966--4985\relax
\mciteBstWouldAddEndPuncttrue
\mciteSetBstMidEndSepPunct{\mcitedefaultmidpunct}
{\mcitedefaultendpunct}{\mcitedefaultseppunct}\relax
\EndOfBibitem
\bibitem[Kasavajjula et~al.(2007)Kasavajjula, Wang, and
  Appleby]{Kasavajjula2007}
Kasavajjula,~U.; Wang,~C.; Appleby,~A.~J. \emph{J. Power Sources}
  \textbf{2007}, \emph{163}, 1003 -- 1039\relax
\mciteBstWouldAddEndPuncttrue
\mciteSetBstMidEndSepPunct{\mcitedefaultmidpunct}
{\mcitedefaultendpunct}{\mcitedefaultseppunct}\relax
\EndOfBibitem
\bibitem[Zhang(2011)]{Zhang2011}
Zhang,~W.-J. \emph{J. Power Sources} \textbf{2011}, \emph{196}, 13 -- 24\relax
\mciteBstWouldAddEndPuncttrue
\mciteSetBstMidEndSepPunct{\mcitedefaultmidpunct}
{\mcitedefaultendpunct}{\mcitedefaultseppunct}\relax
\EndOfBibitem
\bibitem[Szczech and Jin(2011)Szczech, and Jin]{Szczech2011}
Szczech,~J.~R.; Jin,~S. \emph{Energy Environ. Sci.} \textbf{2011}, \emph{4},
  56--72\relax
\mciteBstWouldAddEndPuncttrue
\mciteSetBstMidEndSepPunct{\mcitedefaultmidpunct}
{\mcitedefaultendpunct}{\mcitedefaultseppunct}\relax
\EndOfBibitem
\bibitem[Qian et~al.(2012)Qian, Qiao, Ai, Cao, and Yang]{Qian2012}
Qian,~J.; Qiao,~D.; Ai,~X.; Cao,~Y.; Yang,~H. \emph{Chem. Commun.}
  \textbf{2012}, \emph{48}, 8931--8933\relax
\mciteBstWouldAddEndPuncttrue
\mciteSetBstMidEndSepPunct{\mcitedefaultmidpunct}
{\mcitedefaultendpunct}{\mcitedefaultseppunct}\relax
\EndOfBibitem
\bibitem[Obrovac and Christensen(2004)Obrovac, and Christensen]{Obrovac2004}
Obrovac,~M.~N.; Christensen,~L. \emph{Electrochem. Solid-State Lett.}
  \textbf{2004}, \emph{7}, A93--A96\relax
\mciteBstWouldAddEndPuncttrue
\mciteSetBstMidEndSepPunct{\mcitedefaultmidpunct}
{\mcitedefaultendpunct}{\mcitedefaultseppunct}\relax
\EndOfBibitem
\bibitem[Klein et~al.(2013)Klein, Jache, Bhide, and
  Adelhelm]{klein2013conversion}
Klein,~F.; Jache,~B.; Bhide,~A.; Adelhelm,~P. \emph{Physical Chemistry Chemical
  Physics} \textbf{2013}, \emph{15}, 15876--15887\relax
\mciteBstWouldAddEndPuncttrue
\mciteSetBstMidEndSepPunct{\mcitedefaultmidpunct}
{\mcitedefaultendpunct}{\mcitedefaultseppunct}\relax
\EndOfBibitem
\bibitem[Hong et~al.(2013)Hong, Kim, Park, Choi, Choi, and Lee]{Hong2013}
Hong,~S.~Y.; Kim,~Y.; Park,~Y.; Choi,~A.; Choi,~N.-S.; Lee,~K.~T. \emph{Energy
  Environ. Sci.} \textbf{2013}, \emph{6}, 2067--2081\relax
\mciteBstWouldAddEndPuncttrue
\mciteSetBstMidEndSepPunct{\mcitedefaultmidpunct}
{\mcitedefaultendpunct}{\mcitedefaultseppunct}\relax
\EndOfBibitem
\bibitem[Stevens and Dahn(2001)Stevens, and Dahn]{Stevens2001}
Stevens,~D.~A.; Dahn,~J.~R. \emph{J. Electrochem. Soc.} \textbf{2001},
  \emph{148}, A803--A811\relax
\mciteBstWouldAddEndPuncttrue
\mciteSetBstMidEndSepPunct{\mcitedefaultmidpunct}
{\mcitedefaultendpunct}{\mcitedefaultseppunct}\relax
\EndOfBibitem
\bibitem[Brauer and Zintl(1937)Brauer, and Zintl]{Brauer1937}
Brauer,~G.; Zintl,~E. \emph{Z. Phys. Chem. (Leipzig) Abt. B} \textbf{1937},
  \emph{37(5/6)}, 307--1314\relax
\mciteBstWouldAddEndPuncttrue
\mciteSetBstMidEndSepPunct{\mcitedefaultmidpunct}
{\mcitedefaultendpunct}{\mcitedefaultseppunct}\relax
\EndOfBibitem
\bibitem[Dong and Di~Salvo(2005)Dong, and Di~Salvo]{Dong2005}
Dong,~Y.; Di~Salvo,~F.~J. \emph{Acta Crystallogr. Sect. E: Struct. Rep. Online}
  \textbf{2005}, \emph{61}, i223--i224\relax
\mciteBstWouldAddEndPuncttrue
\mciteSetBstMidEndSepPunct{\mcitedefaultmidpunct}
{\mcitedefaultendpunct}{\mcitedefaultseppunct}\relax
\EndOfBibitem
\bibitem[Li et~al.(2014)Li, Yu, Ye, Ge, Ou, Wu, Feng, Chen, and
  Zhang]{li2014black}
Li,~L.; Yu,~Y.; Ye,~G.~J.; Ge,~Q.; Ou,~X.; Wu,~H.; Feng,~D.; Chen,~X.~H.;
  Zhang,~Y. \emph{Nature nanotechnology} \textbf{2014}, \emph{9},
  372--377\relax
\mciteBstWouldAddEndPuncttrue
\mciteSetBstMidEndSepPunct{\mcitedefaultmidpunct}
{\mcitedefaultendpunct}{\mcitedefaultseppunct}\relax
\EndOfBibitem
\bibitem[Kulish et~al.(2015)Kulish, Malyi, Persson, and Wu]{kulish2015}
Kulish,~V.~V.; Malyi,~O.~I.; Persson,~C.; Wu,~P. \emph{Physical Chemistry
  Chemical Physics} \textbf{2015}, \emph{17}, 13921--13928\relax
\mciteBstWouldAddEndPuncttrue
\mciteSetBstMidEndSepPunct{\mcitedefaultmidpunct}
{\mcitedefaultendpunct}{\mcitedefaultseppunct}\relax
\EndOfBibitem
\bibitem[Zhao et~al.(2014)Zhao, Kang, and Xue]{zhao2014}
Zhao,~S.; Kang,~W.; Xue,~J. \emph{Journal of Materials Chemistry A}
  \textbf{2014}, \emph{2}, 19046--19052\relax
\mciteBstWouldAddEndPuncttrue
\mciteSetBstMidEndSepPunct{\mcitedefaultmidpunct}
{\mcitedefaultendpunct}{\mcitedefaultseppunct}\relax
\EndOfBibitem
\bibitem[Sun et~al.(2015)Sun, Lee, Pasta, Yuan, Zheng, Sun, Li, and
  Cui]{sun2015phosphorene}
Sun,~J.; Lee,~H.-W.; Pasta,~M.; Yuan,~H.; Zheng,~G.; Sun,~Y.; Li,~Y.; Cui,~Y.
  \emph{Nature Nanotechnology} \textbf{2015}, \relax
\mciteBstWouldAddEndPunctfalse
\mciteSetBstMidEndSepPunct{\mcitedefaultmidpunct}
{}{\mcitedefaultseppunct}\relax
\EndOfBibitem
\bibitem[Park and Sohn(2007)Park, and Sohn]{Park2007}
Park,~C.-M.; Sohn,~H.-J. \emph{Adv. Mater.} \textbf{2007}, \emph{19},
  2465--2468\relax
\mciteBstWouldAddEndPuncttrue
\mciteSetBstMidEndSepPunct{\mcitedefaultmidpunct}
{\mcitedefaultendpunct}{\mcitedefaultseppunct}\relax
\EndOfBibitem
\bibitem[Sun et~al.(2012)Sun, Li, Sun, Yu, Wang, and Xie]{Sun2012}
Sun,~L.-Q.; Li,~M.-J.; Sun,~K.; Yu,~S.-H.; Wang,~R.-S.; Xie,~H.-M. \emph{J.
  Phys. Chem. C} \textbf{2012}, \emph{116}, 14772--14779\relax
\mciteBstWouldAddEndPuncttrue
\mciteSetBstMidEndSepPunct{\mcitedefaultmidpunct}
{\mcitedefaultendpunct}{\mcitedefaultseppunct}\relax
\EndOfBibitem
\bibitem[Marino et~al.(2011)Marino, Debenedetti, Fraisse, Favier, and
  Monconduit]{Marino2011}
Marino,~C.; Debenedetti,~A.; Fraisse,~B.; Favier,~F.; Monconduit,~L.
  \emph{Electrochem. Commun.} \textbf{2011}, \emph{13}, 346 -- 349\relax
\mciteBstWouldAddEndPuncttrue
\mciteSetBstMidEndSepPunct{\mcitedefaultmidpunct}
{\mcitedefaultendpunct}{\mcitedefaultseppunct}\relax
\EndOfBibitem
\bibitem[Marino et~al.(2012)Marino, Boulet, Gaveau, Fraisse, and
  Monconduit]{marino2012}
Marino,~C.; Boulet,~L.; Gaveau,~P.; Fraisse,~B.; Monconduit,~L. \emph{Journal
  of Materials Chemistry} \textbf{2012}, \emph{22}, 22713--22720\relax
\mciteBstWouldAddEndPuncttrue
\mciteSetBstMidEndSepPunct{\mcitedefaultmidpunct}
{\mcitedefaultendpunct}{\mcitedefaultseppunct}\relax
\EndOfBibitem
\bibitem[Qian et~al.(2013)Qian, Wu, Cao, Ai, and Yang]{Qian2013}
Qian,~J.; Wu,~X.; Cao,~Y.; Ai,~X.; Yang,~H. \emph{Angew. Chem. Int. Ed.}
  \textbf{2013}, \emph{52}, 4633--4636\relax
\mciteBstWouldAddEndPuncttrue
\mciteSetBstMidEndSepPunct{\mcitedefaultmidpunct}
{\mcitedefaultendpunct}{\mcitedefaultseppunct}\relax
\EndOfBibitem
\bibitem[Ramireddy et~al.(2015)Ramireddy, Xing, Rahman, Chen, Dutercq,
  Gunzelmann, and Glushenkov]{ramireddy2015}
Ramireddy,~T.; Xing,~T.; Rahman,~M.~M.; Chen,~Y.; Dutercq,~Q.; Gunzelmann,~D.;
  Glushenkov,~A.~M. \emph{Journal of Materials Chemistry A} \textbf{2015},
  \emph{3}, 5572--5584\relax
\mciteBstWouldAddEndPuncttrue
\mciteSetBstMidEndSepPunct{\mcitedefaultmidpunct}
{\mcitedefaultendpunct}{\mcitedefaultseppunct}\relax
\EndOfBibitem
\bibitem[Ogata et~al.(2014)Ogata, Salager, Kerr, Fraser, Ducati, Morris,
  Hofmann, and Grey]{Ogata2014}
Ogata,~K.; Salager,~E.; Kerr,~C.~J.; Fraser,~A.~E.; Ducati,~C.; Morris,~A.~J.;
  Hofmann,~S.; Grey,~C.~P. \emph{Nat Commun} \textbf{2014}, \emph{5}\relax
\mciteBstWouldAddEndPuncttrue
\mciteSetBstMidEndSepPunct{\mcitedefaultmidpunct}
{\mcitedefaultendpunct}{\mcitedefaultseppunct}\relax
\EndOfBibitem
\bibitem[Pickard and Needs(2006)Pickard, and Needs]{Pickard2006}
Pickard,~C.~J.; Needs,~R.~J. \emph{Phys. Rev. Lett.} \textbf{2006}, \emph{97},
  045504\relax
\mciteBstWouldAddEndPuncttrue
\mciteSetBstMidEndSepPunct{\mcitedefaultmidpunct}
{\mcitedefaultendpunct}{\mcitedefaultseppunct}\relax
\EndOfBibitem
\bibitem[Pickard and Needs(2011)Pickard, and Needs]{Pickard2011}
Pickard,~C.~J.; Needs,~R.~J. \emph{J. Phys.: Condens. Matter} \textbf{2011},
  \emph{23}, 053201\relax
\mciteBstWouldAddEndPuncttrue
\mciteSetBstMidEndSepPunct{\mcitedefaultmidpunct}
{\mcitedefaultendpunct}{\mcitedefaultseppunct}\relax
\EndOfBibitem
\bibitem[Morris et~al.(2014)Morris, Grey, and Pickard]{Morris2014}
Morris,~A.~J.; Grey,~C.~P.; Pickard,~C.~J. \emph{Phys. Rev. B} \textbf{2014},
  \emph{90}, 054111\relax
\mciteBstWouldAddEndPuncttrue
\mciteSetBstMidEndSepPunct{\mcitedefaultmidpunct}
{\mcitedefaultendpunct}{\mcitedefaultseppunct}\relax
\EndOfBibitem
\bibitem[Jung et~al.(2015)Jung, Allan, Hu, Borkiewicz, Wang, Han, Du, Pickard,
  Chupas, Chapman, Morris, and Grey]{Jung2015(MorrisGe)}
Jung,~H.; Allan,~P.~K.; Hu,~Y.-Y.; Borkiewicz,~O.~J.; Wang,~X.-L.; Han,~W.-Q.;
  Du,~L.-S.; Pickard,~C.~J.; Chupas,~P.~J.; Chapman,~K.~W.; Morris,~A.~J.;
  Grey,~C.~P. \emph{Chemistry of Materials} \textbf{2015}, \emph{27},
  1031--1041\relax
\mciteBstWouldAddEndPuncttrue
\mciteSetBstMidEndSepPunct{\mcitedefaultmidpunct}
{\mcitedefaultendpunct}{\mcitedefaultseppunct}\relax
\EndOfBibitem
\bibitem[See et~al.(2014)See, Leskes, Griffin, Britto, Matthews, Emly, Van~der
  Ven, Wright, Morris, Grey, and Seshadri]{See2014(MorrisS)}
See,~K.~A.; Leskes,~M.; Griffin,~J.~M.; Britto,~S.; Matthews,~P.~D.; Emly,~A.;
  Van~der Ven,~A.; Wright,~D.~S.; Morris,~A.~J.; Grey,~C.~P.; Seshadri,~R.
  \emph{Journal of the American Chemical Society} \textbf{2014}, \emph{136},
  16368--16377, PMID: 25384082\relax
\mciteBstWouldAddEndPuncttrue
\mciteSetBstMidEndSepPunct{\mcitedefaultmidpunct}
{\mcitedefaultendpunct}{\mcitedefaultseppunct}\relax
\EndOfBibitem
\bibitem[Morris et~al.(2008)Morris, Pickard, and Needs]{Morris2008}
Morris,~A.~J.; Pickard,~C.~J.; Needs,~R.~J. \emph{Phys. Rev. B} \textbf{2008},
  \emph{78}, 184102\relax
\mciteBstWouldAddEndPuncttrue
\mciteSetBstMidEndSepPunct{\mcitedefaultmidpunct}
{\mcitedefaultendpunct}{\mcitedefaultseppunct}\relax
\EndOfBibitem
\bibitem[Morris et~al.(2013)Morris, Needs, Salager, Grey, and
  Pickard]{Morris2013}
Morris,~A.~J.; Needs,~R.~J.; Salager,~E.; Grey,~C.~P.; Pickard,~C.~J.
  \emph{Phys. Rev. B} \textbf{2013}, \emph{87}, 174108\relax
\mciteBstWouldAddEndPuncttrue
\mciteSetBstMidEndSepPunct{\mcitedefaultmidpunct}
{\mcitedefaultendpunct}{\mcitedefaultseppunct}\relax
\EndOfBibitem
\bibitem[Clark et~al.(2005)Clark, Segall, Pickard, Hasnip, Probert, Refson, and
  Payne]{CASTEP}
Clark,~S.~J.; Segall,~M.~D.; Pickard,~C.~J.; Hasnip,~P.~J.; Probert,~M.~J.;
  Refson,~K.; Payne,~M. \emph{Z. Kristall.} \textbf{2005}, \emph{220},
  567--570\relax
\mciteBstWouldAddEndPuncttrue
\mciteSetBstMidEndSepPunct{\mcitedefaultmidpunct}
{\mcitedefaultendpunct}{\mcitedefaultseppunct}\relax
\EndOfBibitem
\bibitem[Perdew et~al.(1996)Perdew, Burke, and Ernzerhof]{PBE}
Perdew,~J.~P.; Burke,~K.; Ernzerhof,~M. \emph{Phys. Rev. Lett.} \textbf{1996},
  \emph{77}, 3865--3868\relax
\mciteBstWouldAddEndPuncttrue
\mciteSetBstMidEndSepPunct{\mcitedefaultmidpunct}
{\mcitedefaultendpunct}{\mcitedefaultseppunct}\relax
\EndOfBibitem
\bibitem[Monkhorst and Pack(1976)Monkhorst, and Pack]{MP}
Monkhorst,~H.~J.; Pack,~J.~D. \emph{Phys. Rev. B} \textbf{1976}, \emph{13},
  5188--5192\relax
\mciteBstWouldAddEndPuncttrue
\mciteSetBstMidEndSepPunct{\mcitedefaultmidpunct}
{\mcitedefaultendpunct}{\mcitedefaultseppunct}\relax
\EndOfBibitem
\bibitem[Aydinol et~al.(1997)Aydinol, Kohan, Ceder, Cho, and
  Joannopoulos]{Aydinol1997}
Aydinol,~M.~K.; Kohan,~A.~F.; Ceder,~G.; Cho,~K.; Joannopoulos,~J. \emph{Phys.
  Rev. B} \textbf{1997}, \emph{56}, 1354--1365\relax
\mciteBstWouldAddEndPuncttrue
\mciteSetBstMidEndSepPunct{\mcitedefaultmidpunct}
{\mcitedefaultendpunct}{\mcitedefaultseppunct}\relax
\EndOfBibitem
\bibitem[Pickard and Mauri(2001)Pickard, and Mauri]{GIPAW}
Pickard,~C.; Mauri,~F. \emph{Phys. Rev. B} \textbf{2001}, \emph{63},
  245101\relax
\mciteBstWouldAddEndPuncttrue
\mciteSetBstMidEndSepPunct{\mcitedefaultmidpunct}
{\mcitedefaultendpunct}{\mcitedefaultseppunct}\relax
\EndOfBibitem
\bibitem[Morris et~al.(2014)Morris, Nicholls, Pickard, and Yates]{Morris2014b}
Morris,~A.~J.; Nicholls,~R.~J.; Pickard,~C.~J.; Yates,~J.~R. \emph{Comput.
  Phys. Commun.} \textbf{2014}, \emph{185}, 1477 -- 1485\relax
\mciteBstWouldAddEndPuncttrue
\mciteSetBstMidEndSepPunct{\mcitedefaultmidpunct}
{\mcitedefaultendpunct}{\mcitedefaultseppunct}\relax
\EndOfBibitem
\bibitem[Pickard and Payne(1999)Pickard, and Payne]{Pickard1999}
Pickard,~C.~J.; Payne,~M.~C. \emph{Phys. Rev. B} \textbf{1999}, \emph{59},
  4685--4693\relax
\mciteBstWouldAddEndPuncttrue
\mciteSetBstMidEndSepPunct{\mcitedefaultmidpunct}
{\mcitedefaultendpunct}{\mcitedefaultseppunct}\relax
\EndOfBibitem
\bibitem[Pickard and Payne(2000)Pickard, and Payne]{Pickard2000}
Pickard,~C.~J.; Payne,~M.~C. \emph{Phys. Rev. B} \textbf{2000}, \emph{62},
  4383--4388\relax
\mciteBstWouldAddEndPuncttrue
\mciteSetBstMidEndSepPunct{\mcitedefaultmidpunct}
{\mcitedefaultendpunct}{\mcitedefaultseppunct}\relax
\EndOfBibitem
\bibitem[Refson et~al.(2006)Refson, Tulip, and Clark]{refson2006}
Refson,~K.; Tulip,~P.~R.; Clark,~S.~J. \emph{Physical Review B} \textbf{2006},
  \emph{73}, 155114\relax
\mciteBstWouldAddEndPuncttrue
\mciteSetBstMidEndSepPunct{\mcitedefaultmidpunct}
{\mcitedefaultendpunct}{\mcitedefaultseppunct}\relax
\EndOfBibitem
\bibitem[McNellis et~al.(2009)McNellis, Meyer, and
  Reuter]{mcnellis2009azobenzene}
McNellis,~E.~R.; Meyer,~J.; Reuter,~K. \emph{Physical Review B} \textbf{2009},
  \emph{80}, 205414\relax
\mciteBstWouldAddEndPuncttrue
\mciteSetBstMidEndSepPunct{\mcitedefaultmidpunct}
{\mcitedefaultendpunct}{\mcitedefaultseppunct}\relax
\EndOfBibitem
\bibitem[Grimme(2006)]{grimme2006semiempirical}
Grimme,~S. \emph{Journal of computational chemistry} \textbf{2006}, \emph{27},
  1787--1799\relax
\mciteBstWouldAddEndPuncttrue
\mciteSetBstMidEndSepPunct{\mcitedefaultmidpunct}
{\mcitedefaultendpunct}{\mcitedefaultseppunct}\relax
\EndOfBibitem
\bibitem[Von~Schnering and Wichelhaus(1972)Von~Schnering, and
  Wichelhaus]{Schnering1972}
Von~Schnering,~H.; Wichelhaus,~W. \emph{Naturwissenschaften} \textbf{1972},
  \emph{59}, 78--79\relax
\mciteBstWouldAddEndPuncttrue
\mciteSetBstMidEndSepPunct{\mcitedefaultmidpunct}
{\mcitedefaultendpunct}{\mcitedefaultseppunct}\relax
\EndOfBibitem
\bibitem[Honle et~al.(1983)Honle, Manriquez, Meyer, and von
  Schnering]{Honle1983}
Honle,~W.; Manriquez,~V.; Meyer,~T.; von Schnering,~H. \emph{Z. Kristallogr.}
  \textbf{1983}, \emph{164}, 104--106\relax
\mciteBstWouldAddEndPuncttrue
\mciteSetBstMidEndSepPunct{\mcitedefaultmidpunct}
{\mcitedefaultendpunct}{\mcitedefaultseppunct}\relax
\EndOfBibitem
\bibitem[Honle and von Schnering(1981)Honle, and von Schnering]{Honle1981}
Honle,~W.; von Schnering,~H. \emph{Z. Kristallogr.} \textbf{1981}, \emph{155},
  307--1314\relax
\mciteBstWouldAddEndPuncttrue
\mciteSetBstMidEndSepPunct{\mcitedefaultmidpunct}
{\mcitedefaultendpunct}{\mcitedefaultseppunct}\relax
\EndOfBibitem
\bibitem[Dong and DiSalvo(2007)Dong, and DiSalvo]{Dong2007}
Dong,~Y.; DiSalvo,~F.~J. \emph{Acta Crystallographica Section E} \textbf{2007},
  \emph{63}, i97--i98\relax
\mciteBstWouldAddEndPuncttrue
\mciteSetBstMidEndSepPunct{\mcitedefaultmidpunct}
{\mcitedefaultendpunct}{\mcitedefaultseppunct}\relax
\EndOfBibitem
\bibitem[Wichelhaus and von Schnering(1973)Wichelhaus, and von
  Schnering]{Wichelhaus1973}
Wichelhaus,~W.; von Schnering,~H. \emph{Naturwissenschaften} \textbf{1973},
  \emph{60}, 104--104\relax
\mciteBstWouldAddEndPuncttrue
\mciteSetBstMidEndSepPunct{\mcitedefaultmidpunct}
{\mcitedefaultendpunct}{\mcitedefaultseppunct}\relax
\EndOfBibitem
\bibitem[Ozisik et~al.(2011)Ozisik, Colakoglu, Deligoz, and Ozisik]{Ozisik2011}
Ozisik,~H.; Colakoglu,~K.; Deligoz,~E.; Ozisik,~H. \emph{Solid State
  Communications} \textbf{2011}, \emph{151}, 1349 -- 1354\relax
\mciteBstWouldAddEndPuncttrue
\mciteSetBstMidEndSepPunct{\mcitedefaultmidpunct}
{\mcitedefaultendpunct}{\mcitedefaultseppunct}\relax
\EndOfBibitem
\bibitem[Ivanov et~al.(2012)Ivanov, Morris, Bozhenko, Pickard, and
  Boldyrev]{Ivanov2012}
Ivanov,~A.~S.; Morris,~A.~J.; Bozhenko,~K.~V.; Pickard,~C.~J.; Boldyrev,~A.~I.
  \emph{Angew. Chem. Int. Ed.} \textbf{2012}, \emph{51}, 8330--8333\relax
\mciteBstWouldAddEndPuncttrue
\mciteSetBstMidEndSepPunct{\mcitedefaultmidpunct}
{\mcitedefaultendpunct}{\mcitedefaultseppunct}\relax
\EndOfBibitem
\bibitem[Jung et~al.(2015)Jung, Allan, Hu, Borkiewicz, Wang, Han, Du, Pickard,
  Chupas, Chapman, Morris, and Grey]{Hyeyoung20015}
Jung,~H.; Allan,~P.~K.; Hu,~Y.-Y.; Borkiewicz,~O.~J.; Wang,~X.-L.; Han,~W.-Q.;
  Du,~L.-S.; Pickard,~C.~J.; Chupas,~P.~J.; Chapman,~K.~W.; Morris,~A.~J.;
  Grey,~C.~P. \emph{Chemistry of Materials} \textbf{2015}, \emph{27},
  1031--1041\relax
\mciteBstWouldAddEndPuncttrue
\mciteSetBstMidEndSepPunct{\mcitedefaultmidpunct}
{\mcitedefaultendpunct}{\mcitedefaultseppunct}\relax
\EndOfBibitem
\bibitem[auf~der G{\"{u}}nne et~al.(1999)auf~der G{\"{u}}nne, Kaczmarek, van
  W{\"{u}}llen, Eckert, Paschke, Foecker, and Jeitschko]{Gunne1999}
auf~der G{\"{u}}nne,~J.~S.; Kaczmarek,~S.; van W{\"{u}}llen,~L.; Eckert,~H.;
  Paschke,~D.; Foecker,~A.~J.; Jeitschko,~W. \emph{Journal of Solid State
  Chemistry} \textbf{1999}, \emph{147}, 341 -- 349\relax
\mciteBstWouldAddEndPuncttrue
\mciteSetBstMidEndSepPunct{\mcitedefaultmidpunct}
{\mcitedefaultendpunct}{\mcitedefaultseppunct}\relax
\EndOfBibitem
\bibitem[Laskowski et~al.(2013)Laskowski, Blaha, and Tran]{laskowski2013}
Laskowski,~R.; Blaha,~P.; Tran,~F. \emph{Physical Review B} \textbf{2013},
  \emph{87}, 195130\relax
\mciteBstWouldAddEndPuncttrue
\mciteSetBstMidEndSepPunct{\mcitedefaultmidpunct}
{\mcitedefaultendpunct}{\mcitedefaultseppunct}\relax
\EndOfBibitem
\bibitem[Rehr et~al.(1995)Rehr, Guerra, Parkin, Hope, and Kauzlarich]{Rehr1995}
Rehr,~A.; Guerra,~F.; Parkin,~S.; Hope,~H.; Kauzlarich,~S.~M. \emph{Inorganic
  Chemistry} \textbf{1995}, \emph{34}, 6218--6220\relax
\mciteBstWouldAddEndPuncttrue
\mciteSetBstMidEndSepPunct{\mcitedefaultmidpunct}
{\mcitedefaultendpunct}{\mcitedefaultseppunct}\relax
\EndOfBibitem
\bibitem[Abicht et~al.(1984)Abicht, H~nle, and v.~Schnering]{Abicht1984}
Abicht,~H.-P.; H~nle,~W.; v.~Schnering,~H.~G. \emph{Zeitschrift fur
  anorganische und allgemeine Chemie} \textbf{1984}, \emph{519}, 7--23\relax
\mciteBstWouldAddEndPuncttrue
\mciteSetBstMidEndSepPunct{\mcitedefaultmidpunct}
{\mcitedefaultendpunct}{\mcitedefaultseppunct}\relax
\EndOfBibitem
\bibitem[von Schnering et~al.(1989)von Schnering, Hartweg, Hartweg, and
  Honle]{Schnering1989}
von Schnering,~H.~G.; Hartweg,~M.; Hartweg,~U.; Honle,~W. \emph{Angewandte
  Chemie} \textbf{1989}, \emph{101}, 98--99\relax
\mciteBstWouldAddEndPuncttrue
\mciteSetBstMidEndSepPunct{\mcitedefaultmidpunct}
{\mcitedefaultendpunct}{\mcitedefaultseppunct}\relax
\EndOfBibitem
\bibitem[Chen and Yamanaka(2004)Chen, and Yamanaka]{Chen2004}
Chen,~X.; Yamanaka,~S. \emph{J. Alloys Compd.} \textbf{2004}, \emph{370}, 110
  -- 113\relax
\mciteBstWouldAddEndPuncttrue
\mciteSetBstMidEndSepPunct{\mcitedefaultmidpunct}
{\mcitedefaultendpunct}{\mcitedefaultseppunct}\relax
\EndOfBibitem
\end{mcitethebibliography}

\end{document}